\input harvmac
\input epsf
\input amssym
%
%
\noblackbox
\newcount\figno
\figno=0
\def\fig#1#2#3{
\par\begingroup\parindent=0pt\leftskip=1cm\rightskip=1cm\parindent=0pt
\baselineskip=11pt
\global\advance\figno by 1
\midinsert
\epsfxsize=#3
\centerline{\epsfbox{#2}}
\vskip -21pt
{\bf Fig.\ \the\figno: } #1\par
\endinsert\endgroup\par
}
\def\figlabel#1{\xdef#1{\the\figno}}
\def\encadremath#1{\vbox{\hrule\hbox{\vrule\kern8pt\vbox{\kern8pt
\hbox{$\displaystyle #1$}\kern8pt}
\kern8pt\vrule}\hrule}}

\def\frac#1#2{{#1 \over #2}}

\def\p{\partial}
\def\semi{\subset\kern-1em\times\;}

\def\sqr#1#2{{\vcenter{\vbox{\hrule height.#2pt
\hbox{\vrule width.#2pt height#1pt \kern#1pt \vrule width.#2pt}
\hrule height.#2pt}}}}

\def\p{\partial}

\def\th{\theta}

\def\thetah{\hat{\theta}}
\def\phih{\hat{\phi}}

\def\rh{\hat{r}}
\def\th{\hat{t}}

\def\p{\partial}

%

%

\def\IR{\Bbb{R}}

\def\zb{\overline{z}}
\def\th{\hat{t}}

\def\Nc{{\cal N}}
\def\qh{\hat{q}}
\def\Mh{\hat{M}}


\lref\masakiref{
  N.~Iizuka and M.~Shigemori,
  ``A Note on D1-D5-J System and 5D Small Black Ring,''
  arXiv:hep-th/0506215.
}

\lref\KLatt{
  P.~Kraus and F.~Larsen,
  ``Attractors and black rings,''
  Phys.\ Rev.\ D {\bf 72}, 024010 (2005)
  [arXiv:hep-th/0503219].
}

\lref\ChamseddinePI{
  A.~H.~Chamseddine, S.~Ferrara, G.~W.~Gibbons and R.~Kallosh,
  ``Enhancement of supersymmetry near 5d black hole horizon,''
  Phys.\ Rev.\ D {\bf 55}, 3647 (1997)
  [arXiv:hep-th/9610155].
}

\lref\StromBTZ{ A.~Strominger,
 ``Black hole entropy from near-horizon microstates'',
JHEP {\bf 9802}, 009 (1998); [arXiv:hep-th/9712251];
V.~Balasubramanian and F.~Larsen, ``Near horizon geometry and black
holes in four dimensions'', Nucl.\ Phys.\ B {\bf 528}, 229 (1998);
[arXiv:hep-th/9802198].
}

\lref\BalasubramanianEE{
  V.~Balasubramanian and F.~Larsen,
  Nucl.\ Phys.\  B {\bf 528}, 229 (1998)
  [arXiv:hep-th/9802198].
}

\lref\MSW{ J.~M.~Maldacena, A.~Strominger and E.~Witten, ``Black
hole entropy in M-theory'', JHEP {\bf 9712}, 002 (1997);
[arXiv:hep-th/9711053]. }

\lref\HMM{ J.~A.~Harvey, R.~Minasian and G.~W.~Moore, ``Non-abelian
tensor-multiplet anomalies,''
 JHEP {\bf 9809}, 004 (1998)
  [arXiv:hep-th/9808060].
}

\lref\antRRRR{  I.~Antoniadis, S.~Ferrara, R.~Minasian and
K.~S.~Narain,
  ``R**4 couplings in M- and type II theories on Calabi-Yau spaces,''
  Nucl.\ Phys.\ B {\bf 507}, 571 (1997)
  [arXiv:hep-th/9707013].
 }

\lref\WittenMfive{ E.~Witten,
  ``Five-brane effective action in M-theory,''
  J.\ Geom.\ Phys.\  {\bf 22}, 103 (1997)
  [arXiv:hep-th/9610234].
}

\lref\wittenAdS{ E.~Witten,
  ``Anti-de Sitter space and holography,''
  Adv.\ Theor.\ Math.\ Phys.\  {\bf 2}, 253 (1998)
  [arXiv:hep-th/9802150].
  }

\lref\brownhen{  J.~D.~Brown and M.~Henneaux,
 ``Central Charges In The Canonical Realization Of Asymptotic Symmetries: An
  Example From Three-Dimensional Gravity,''
  Commun.\ Math.\ Phys.\  {\bf 104}, 207 (1986).
  }

\lref\wald{
  R.~M.~Wald,
  ``Black hole entropy is the Noether charge,''
  Phys.\ Rev.\ D {\bf 48}, 3427 (1993)
  [arXiv:gr-qc/9307038].
R.~Wald, Phys.\ Rev.\ D {\bf 48} R3427 (1993);
   V.~Iyer and R.~M.~Wald,
  ``Some properties of Noether charge and a proposal for dynamical black hole
  entropy,''
  Phys.\ Rev.\ D {\bf 50}, 846 (1994)
  [arXiv:gr-qc/9403028].
 ``A Comparison of Noether charge and Euclidean methods for computing the
  entropy of stationary black holes,''
  Phys.\ Rev.\ D {\bf 52}, 4430 (1995)
  [arXiv:gr-qc/9503052].
}

\lref\senrescaled{A.~Sen,
   ``Black holes, elementary strings and holomorphic anomaly,''
     JHEP {\bf 0507}, 063 (2005)
  [arXiv:hep-th/0502126];
   ``Entropy function for heterotic black holes,''
      JHEP {\bf 0603}, 008 (2006)
  [arXiv:hep-th/0508042]
;
  B.~Sahoo and A.~Sen,
  ``alpha' corrections to extremal dyonic black holes in heterotic string
  theory,''
  JHEP {\bf 0701}, 010 (2007)
  [arXiv:hep-th/0608182].
}

\lref\SenWA{
  A.~Sen,
  ``Black hole entropy function and the attractor mechanism in higher
  derivative gravity,''
  JHEP {\bf 0509}, 038 (2005)
  [arXiv:hep-th/0506177].
}

\lref\saidasoda{
  H.~Saida and J.~Soda,
  ``Statistical entropy of BTZ black hole in higher curvature gravity,''
  Phys.\ Lett.\ B {\bf 471}, 358 (2000)
  [arXiv:gr-qc/9909061].
}

\lref\attract{ S.~Ferrara, R.~Kallosh and A.~Strominger, ``N=2
extremal black holes'', Phys.\ Rev.\ D {\bf 52}, 5412 (1995),
[arXiv:hep-th/9508072];
 A.~Strominger,
 ``Macroscopic Entropy of $N=2$ Extremal Black Holes'',
 Phys.\ Lett.\ B {\bf 383}, 39 (1996),
[arXiv:hep-th/9602111];
S.~Ferrara and R.~Kallosh, ``Supersymmetry and Attractors'', Phys.\
Rev.\ D {\bf 54}, 1514 (1996), [arXiv:hep-th/9602136];
``Universality of Supersymmetric Attractors'', Phys.\ Rev.\ D {\bf
54}, 1525 (1996), [arXiv:hep-th/9603090];
R.~Kallosh, A.~Rajaraman and W.~K.~Wong, ``Supersymmetric rotating
black holes and attractors'', Phys.\ Rev.\ D {\bf 55}, 3246 (1997),
[arXiv:hep-th/9611094];
A~Chou, R.~Kallosh, J.~Rahmfeld, S.~J.~Rey, M.~Shmakova and
W.~K.~Wong, ``Critical points and phase transitions in 5d
compactifications of M-theory''. Nucl.\ Phys.\ B {\bf 508}, 147
(1997); [arXiv:hep-th/9704142].
}

\lref\moore{G.~W.~Moore,``Attractors and arithmetic'',
[arXiv:hep-th/9807056];
``Arithmetic and attractors'', [arXiv:hep-th/9807087];
``Les Houches lectures on strings and arithmetic'',
[arXiv:hep-th/0401049];
B.~R.~Greene and C.~I.~Lazaroiu, ``Collapsing D-branes in Calabi-Yau
moduli space. I'', Nucl.\ Phys.\ B {\bf 604}, 181 (2001),
[arXiv:hep-th/0001025]. }

\lref\ChamseddinePI{
  A.~H.~Chamseddine, S.~Ferrara, G.~W.~Gibbons and R.~Kallosh,
  ``Enhancement of supersymmetry near 5d black hole horizon,''
  Phys.\ Rev.\ D {\bf 55}, 3647 (1997)
  [arXiv:hep-th/9610155].
}

\lref\denef{  
F.~Denef,``Supergravity flows and D-brane stability'', JHEP {\bf
0008}, 050 (2000), [arXiv:hep-th/0005049];
``On the correspondence between D-branes and stationary supergravity
 solutions of type II Calabi-Yau compactifications'',
[arXiv:hep-th/0010222];
``(Dis)assembling special Lagrangians'', [arXiv:hep-th/0107152].
  B.~Bates and F.~Denef,
   ``Exact solutions for supersymmetric stationary black hole composites,''
  arXiv:hep-th/0304094.
}

\lref\OSV{H.~Ooguri, A.~Strominger and C.~Vafa, ``Black hole
attractors and the topological string'', Phys.\ Rev.\ D {\bf 70},
106007 (2004), [arXiv:hep-th/0405146];
}

\lref\moreOSV{ J.~de Boer, M.~C.~N.~Cheng, R.~Dijkgraaf, J.~Manschot
and E.~Verlinde, ``A farey tail for attractor black holes,'' JHEP
{\bf 0611}, 024 (2006) [arXiv:hep-th/0608059].
}

\lref\DabholkarYR{
  A.~Dabholkar,
  ``Exact counting of black hole microstates,''
  Phys.\ Rev.\ Lett.\  {\bf 94}, 241301 (2005)
  [arXiv:hep-th/0409148].
}

 \lref\DDMP{
A.~Dabholkar, F.~Denef, G.~W.~Moore and B.~Pioline, ``Exact and
asymptotic degeneracies of small black holes'',
[arXiv:hep-th/0502157];  ``Precision counting of small black
holes,''
  JHEP {\bf 0510}, 096 (2005)
  [arXiv:hep-th/0507014].
}

\lref\curvcorr{A.~Dabholkar, ``Exact counting of black hole
microstates", [arXiv:hep-th/0409148],
A.~Dabholkar, R.~Kallosh and A.~Maloney, ``A stringy cloak for a
classical singularity'', JHEP {\bf 0412}, 059 (2004),
[arXiv:hep-th/0410076].
} \lref\bkmicro{
 I.~Bena and P.~Kraus,
 ``Microscopic description of black rings in AdS/CFT'',
JHEP {\bf 0412}, 070 (2004)
  [arXiv:hep-th/0408186].
} \lref\cgms{ M.~Cyrier, M.~Guica, D.~Mateos and A.~Strominger,
``Microscopic entropy of the black ring'', [arXiv:hep-th/0411187].
}

\lref\CardosoFP{
  K.~Behrndt, G.~Lopes Cardoso, B.~de Wit, D.~Lust, T.~Mohaupt and
  W.~A.~Sabra,
  ``Higher-order black-hole solutions in N = 2 supergravity and
  Calabi-Yau
  string backgrounds,''
  Phys.\ Lett.\ B {\bf 429}, 289 (1998)
  [arXiv:hep-th/9801081];G.~Lopes Cardoso, B.~de Wit, D.~Lust,
  T.~Mohaupt,
  ``Corrections to macroscopic supersymmetric black-hole entropy'',
  Phys.\ Lett.\ B {\bf 451}, 309 (1999)
  [arXiv:hep-th/9812082].
    ``Macroscopic entropy formulae and non-holomorphic corrections for
  supersymmetric black holes'',
  Nucl.\ Phys.\ B {\bf 567}, 87 (2000)
  [arXiv:hep-th/9906094];
G.~Lopes Cardoso, B.~de Wit, J.~Kappeli , T.~Mohaupt
 ``Stationary BPS solutions in N = 2
supergravity with $R^2 $ interactions'', JHEP {\bf 0012}, 019 (2000)
[arXiv:hep-th/0009234];
  }

\lref\hensken{  M.~Henningson and K.~Skenderis,
  ``The holographic Weyl anomaly,''
  JHEP {\bf 9807}, 023 (1998)
  [arXiv:hep-th/9806087].
  }

\lref\balkraus{  V.~Balasubramanian and P.~Kraus,
  ``A stress tensor for anti-de Sitter gravity,''
  Commun.\ Math.\ Phys.\  {\bf 208}, 413 (1999)
  [arXiv:hep-th/9902121].
  }

\lref\HanakiPJ{
  K.~Hanaki, K.~Ohashi and Y.~Tachikawa,
  ``Supersymmetric completion of an R**2 term in five-dimensional
  supergravity,''
  [arXiv:hep-th/0611329].
}

\lref\us{
  P.~Kraus and F.~Larsen,
  ``Microscopic black hole entropy in theories with higher derivatives,''
  JHEP {\bf 0509}, 034 (2005)
  [arXiv:hep-th/0506176].
}

\lref\SahooVZ{
  B.~Sahoo and A.~Sen,
  ``BTZ black hole with Chern-Simons and higher derivative terms,''
  JHEP {\bf 0607}, 008 (2006)
  [arXiv:hep-th/0601228].
}

\lref\GutowskiYV{
  J.~B.~Gutowski and H.~S.~Reall,
  ``General supersymmetric AdS(5) black holes,''
  JHEP {\bf 0404}, 048 (2004)
  [arXiv:hep-th/0401129].
}

\lref\LarsenXM{
  F.~Larsen,
  ``The attractor mechanism in five dimensions,''
  [arXiv:hep-th/0608191].
}

\lref\MaldacenaBW{
  J.~M.~Maldacena and A.~Strominger,
  ``AdS(3) black holes and a stringy exclusion principle,''
  JHEP {\bf 9812}, 005 (1998)
  [arXiv:hep-th/9804085].
}

\lref\BanadosGQ{
  M.~Banados, M.~Henneaux, C.~Teitelboim and J.~Zanelli,
  ``Geometry of the (2+1) black hole,''
  Phys.Rev.D {\bf 66}, 010001 (2002)
  [arXiv:gr-qc/9302012].
}

\lref\StromingerYG{
  A.~Strominger,
  ``AdS(2) quantum gravity and string theory,''
  JHEP {\bf 9901}, 007 (1999)
  [arXiv:hep-th/9809027].
}

\lref\MohauptMJ{
  T.~Mohaupt,
  ``Black hole entropy, special geometry and strings,''
  Phys.Rev.D {\bf 66}, 010001 (2002)
  arXiv:hep-th/0007195;
  ``Supersymmetric black holes in string theory,''
  arXiv:hep-th/0703035.
}

\lref\BehrndtHE{
  K.~Behrndt, G.~Lopes Cardoso and S.~Mahapatra,
  ``Exploring the relation between 4D and 5D BPS solutions,''
  Nucl.\ Phys.\  B {\bf 732}, 200 (2006)
  [arXiv:hep-th/0506251].
}

\lref\BenaNI{
  I.~Bena, P.~Kraus and N.~P.~Warner,
  ``Black rings in Taub-NUT,''
  Phys.\ Rev.\  D {\bf 72}, 084019 (2005)
  [arXiv:hep-th/0504142].
}

\lref\GaiottoGF{
  D.~Gaiotto, A.~Strominger and X.~Yin,
  ``New connections between 4D and 5D black holes,''
  JHEP {\bf 0602}, 024 (2006)
  [arXiv:hep-th/0503217].
}

\lref\GaiottoXT{
  D.~Gaiotto, A.~Strominger and X.~Yin,
  ``5D black rings and 4D black holes,''
  JHEP {\bf 0602}, 023 (2006)
  [arXiv:hep-th/0504126].
}

\lref\GuicaIG{
  M.~Guica, L.~Huang, W.~Li and A.~Strominger,
  ``R**2 corrections for 5D black holes and rings,''
  JHEP {\bf 0610}, 036 (2006)
  [arXiv:hep-th/0505188].
}

\lref\KrausZM{
  P.~Kraus and F.~Larsen,
  ``Holographic gravitational anomalies,''
  JHEP {\bf 0601}, 022 (2006)
  [arXiv:hep-th/0508218].
}

\lref\GaiottoNS{
  D.~Gaiotto, A.~Strominger and X.~Yin,
  ``From AdS(3)/CFT(2) to black holes / topological strings,''
  [arXiv:hep-th/0602046].
}

\lref\MohauptMJ{
  T.~Mohaupt,
  ``Black hole entropy, special geometry and strings,''
  Fortsch.\ Phys.\  {\bf 49}, 3 (2001)
  [arXiv:hep-th/0007195].
}

\lref\PiolineNI{
  B.~Pioline,
  ``Lectures on black holes, topological strings and quantum attractors,''
  Class.\ Quant.\ Grav.\  {\bf 23}, S981 (2006)
  [arXiv:hep-th/0607227].
}

\lref\KrausWN{
  P.~Kraus,
  ``Lectures on black holes and the AdS(3)/CFT(2) correspondence,''
  [arXiv:hep-th/0609074].
}

\lref\DabholkarTB{
  A.~Dabholkar, A.~Sen and S.~P.~Trivedi,
  ``Black hole microstates and attractor without supersymmetry,''
  JHEP {\bf 0701}, 096 (2007)
  [arXiv:hep-th/0611143].
}

\lref\rings{
  I.~Bena and P.~Kraus,
  ``R**2 corrections to black ring entropy,''
  arXiv:hep-th/0506015;
  N.~Iizuka and M.~Shigemori,
  ``A note on D1-D5-J system and 5D small black ring,''
  JHEP {\bf 0508}, 100 (2005)
  [arXiv:hep-th/0506215]; A.~Dabholkar, N.~Iizuka, A.~Iqubal and M.~Shigemori,
  ``Precision microstate counting of small black rings,''
  Phys.\ Rev.\ Lett.\  {\bf 96}, 071601 (2006)
  [arXiv:hep-th/0511120]; A.~Dabholkar, N.~Iizuka, A.~Iqubal, A.~Sen and M.~Shigemori,
  ``Spinning strings as small black rings,''
 [arXiv:hep-th/0611166].
}

\lref\TachikawaSZ{
  Y.~Tachikawa,
  ``Black hole entropy in the presence of Chern-Simons terms,''
  Class.\ Quant.\ Grav.\  {\bf 24}, 737 (2007)
  [arXiv:hep-th/0611141].
}

\lref\KugoHN{
  T.~Kugo and K.~Ohashi,
  ``Supergravity tensor calculus in 5D from 6D,''
  Prog.\ Theor.\ Phys.\  {\bf 104}, 835 (2000)
  [arXiv:hep-ph/0006231]; T.~Fujita and K.~Ohashi,
  ``Superconformal tensor calculus in five dimensions,''
  Prog.\ Theor.\ Phys.\  {\bf 106}, 221 (2001)
  [arXiv:hep-th/0104130].
}

\lref\BergshoeffHC{
  E.~Bergshoeff, S.~Cucu, M.~Derix, T.~de Wit, R.~Halbersma and A.~Van Proeyen,
  ``Weyl multiplets of N = 2 conformal supergravity in five dimensions,''
  JHEP {\bf 0106}, 051 (2001)
  [arXiv:hep-th/0104113].
  ``N = 2 supergravity in five dimensions revisited,''
  Class.\ Quant.\ Grav.\  {\bf 21}, 3015 (2004)
  [Class.\ Quant.\ Grav.\  {\bf 23}, 7149 (2006)]
  [arXiv:hep-th/0403045].
}

\lref\DabholkarDQ{
  A.~Dabholkar, R.~Kallosh and A.~Maloney,
  ``A stringy cloak for a classical singularity,''
  JHEP {\bf 0412}, 059 (2004)
  [arXiv:hep-th/0410076];
  V.~Hubeny, A.~Maloney and M.~Rangamani,
  ``String-corrected black holes,''
  JHEP {\bf 0505}, 035 (2005)
  [arXiv:hep-th/0411272].
}

\lref\senstretch{A.~Sen,
  ``How does a fundamental string stretch its horizon?,''
  JHEP {\bf 0505}, 059 (2005)
  [arXiv:hep-th/0411255];
  }

\lref\KrausNB{
  P.~Kraus and F.~Larsen,
  ``Partition functions and elliptic genera from supergravity,''
  JHEP {\bf 0701}, 002 (2007)
  [arXiv:hep-th/0607138].
}

\lref\CastroSD{
  A.~Castro, J.~L.~Davis, P.~Kraus and F.~Larsen,
  ``5D attractors with higher derivatives,''
  arXiv:hep-th/0702072.
}

\lref\moreCDKL{
  A.~Castro, J.~L.~Davis, P.~Kraus and F.~Larsen.
  In progress.
}

\lref\CadavidBK{
  A.~C.~Cadavid, A.~Ceresole, R.~D'Auria and S.~Ferrara,
  Phys.\ Lett.\  B {\bf 357}, 76 (1995)
  [arXiv:hep-th/9506144].
}

\lref\ShmakovaNZ{
  M.~Shmakova,
  ``Calabi-Yau black holes,''
  Phys.\ Rev.\  D {\bf 56}, 540 (1997)
  [arXiv:hep-th/9612076].
}

\lref\ChouBA{
  A.~Chou, R.~Kallosh, J.~Rahmfeld, S.~J.~Rey, M.~Shmakova and W.~K.~Wong,
  ``Critical points and phase transitions in 5d compactifications of
  M-theory,''
  Nucl.\ Phys.\  B {\bf 508}, 147 (1997)
  [arXiv:hep-th/9704142].
}

\lref\MaldacenaDE{
  J.~M.~Maldacena, A.~Strominger and E.~Witten,
  ``Black hole entropy in M-theory,''
  JHEP {\bf 9712}, 002 (1997)
  [arXiv:hep-th/9711053].
}

\lref\VafaGR{
  C.~Vafa,
  ``Black holes and Calabi-Yau threefolds,''
  Adv.\ Theor.\ Math.\ Phys.\  {\bf 2}, 207 (1998)
  [arXiv:hep-th/9711067].
}

\lref\HarveyBX{
  J.~A.~Harvey, R.~Minasian and G.~W.~Moore,
  ``Non-abelian tensor-multiplet anomalies,''
  JHEP {\bf 9809}, 004 (1998)
  [arXiv:hep-th/9808060].
}

\lref\BenaAY{
  I.~Bena and P.~Kraus,
  `Microstates of the D1-D5-KK system,''
  Phys.\ Rev.\  D {\bf 72}, 025007 (2005)
  [arXiv:hep-th/0503053]; ``Microscopic description of black rings in AdS/CFT,''
  JHEP {\bf 0412}, 070 (2004)
  [arXiv:hep-th/0408186].
}

\lref\ElvangSA{
  H.~Elvang, R.~Emparan, D.~Mateos and H.~S.~Reall,
  ``Supersymmetric 4D rotating black holes from 5D black rings,''
  JHEP {\bf 0508}, 042 (2005)
  [arXiv:hep-th/0504125].
}

\lref\StromingerSH{
  A.~Strominger and C.~Vafa,
  ``Microscopic Origin of the Bekenstein-Hawking Entropy,''
  Phys.\ Lett.\  B {\bf 379}, 99 (1996)
  [arXiv:hep-th/9601029].
}

\lref\BreckenridgeIS{
  J.~C.~Breckenridge, R.~C.~Myers, A.~W.~Peet and C.~Vafa,
  ``D-branes and spinning black holes,''
  Phys.\ Lett.\  B {\bf 391}, 93 (1997)
  [arXiv:hep-th/9602065].
}

\lref\GiveonPR{
  A.~Giveon and D.~Kutasov,
  ``Fundamental strings and black holes,''
  JHEP {\bf 0701}, 071 (2007)
  [arXiv:hep-th/0611062].
}

\lref\ZwiebachUQ{
  B.~Zwiebach,
  ``Curvature Squared Terms And String Theories,''
  Phys.\ Lett.\  B {\bf 156}, 315 (1985).
}

\lref\ChamseddineQS{
  A.~H.~Chamseddine and W.~A.~Sabra,
  ``Calabi-Yau black holes and enhancement of supersymmetry in five
  dimensions,''
  Phys.\ Lett.\  B {\bf 460}, 63 (1999)
  [arXiv:hep-th/9903046].
}

\lref\SabraYD{
  W.~A.~Sabra,
  Mod.\ Phys.\ Lett.\  A {\bf 13}, 239 (1998)
  [arXiv:hep-th/9708103].
}

\lref\CastroCI{
  A.~Castro, J.~L.~Davis, P.~Kraus and F.~Larsen,
  ``Precision entropy of spinning black holes,''
  arXiv:0705.1847 [hep-th].
}


\Title{\vbox{\baselineskip12pt
}} {\vbox{\centerline {5D Black Holes and Strings with Higher Derivatives}}}
\centerline{Alejandra
Castro$^\dagger$\foot{aycastro@umich.edu}, Joshua L.
Davis$^{\spadesuit}$\foot{davis@physics.ucla.edu}, Per
Kraus$^{\spadesuit}$\foot{pkraus@ucla.edu}, and Finn
Larsen$^\dagger$\foot{larsenf@umich.edu}}
\bigskip
\centerline{${}^\dagger$\it{Department of Physics
and Michigan Center for Theoretical Physics,
}} \centerline{\it{University of Michigan, Ann
Arbor, MI 48109-1120, USA.}}\vskip.2cm
\centerline{${}^{\spadesuit}$\it{Department of Physics and
Astronomy, UCLA,}}\centerline{\it{ Los Angeles, CA 90095-1547,
USA.}}

\baselineskip15pt

\vskip .3in

\centerline{\bf Abstract}

We find asymptotically flat black hole and string solutions to 5D
supergravity in the presence of higher derivative terms. In some
cases, including the fundamental heterotic string solution, the
higher derivative terms smooth out naked singularities into
regular geometries carrying zero entropy. We also compute
corrections to the entropy of 5D Calabi-Yau black holes, and
discuss the relation to previous results.

\Date{March, 2007}
\baselineskip14pt

\newsec{Introduction}

In this paper we find asymptotically flat solutions to 5D
supergravity in the presence of four derivative terms. We obtain
two types of solutions: magnetic string solutions, with near
horizon geometry AdS$_3 \times S^2$; and electric particle
solutions, with near horizon geometry AdS$_2 \times S^3$.   These
spacetimes are central to the study of black holes in string
theory.  Namely, the electric solutions represent 5D black holes,
while the magnetic solutions yield 4D black holes upon
compactification on a circle with momentum added.

The inclusion of higher derivative $R^2$ terms allows us to
compute corrections to quantities such as the black hole entropy,
and also to obtain smooth spacetimes in cases where the two
derivative action  yields a naked singularity.  Via the AdS/CFT
correspondence, this permits a more detailed comparison between
the microscopic and macroscopic descriptions of black holes in
string theory.

The program of including $R^2$ corrections in  4D supergravity has
received much attention over the past few years
\refs{\CardosoFP,\OSV,\senrescaled,\DabholkarYR,\DDMP,\us,\SenWA}
(for reviews see \refs{\MohauptMJ,\PiolineNI,\KrausWN}). By
contrast, the relevant 5D supergravity action was constructed only
very recently, as  the supersymmetric completion of a certain
mixed gauge/gravitational Chern-Simons term \HanakiPJ. The 5D
theory can be thought of as arising from M-theory compactified on
a Calabi-Yau threefold. The relevant higher derivative terms are
determined by a combination of anomalies and supersymmetry.

Before proceeding, let us note some of the general advantages of
working in a 5D setting. First, by compactifying one of our 5D
directions on a circle (or more generally, a Taub-NUT fiber) we
can reproduce all of the 4D solutions of interest
\refs{\BenaAY,\ElvangSA,\GaiottoGF,\GaiottoXT,\BenaNI,\BehrndtHE}.
On the other hand, by not compactifying we maintain access to
solutions that are inherently five dimensional. One example is the
standard  5D black hole \refs{\StromingerSH,\BreckenridgeIS},
which we'll refer to as the electric black hole. Another example
is a straight fundamental heterotic string with zero momentum,
which provides the simplest example of our magnetic solutions. To
access these in the 4D description requires the auxiliary
procedure of decompactifying a Taub-NUT fiber; here the
description is simpler and more direct.

In this paper we restrict attention to spherically symmetric
solutions to simplify the analysis.  The generalization to
spinning black holes and black rings  will appear in subsequent
work \moreCDKL. As we'll see, a very nice feature of the off-shell
$R^2$ supergravity obtained in \HanakiPJ\ is that the construction
of BPS solutions is surprisingly simple.  Indeed, the bulk of the
analysis is entirely parallel to the two derivative case. The
higher derivatives only manifest themselves towards the end of the
construction, where they yield corrections to the standard special
geometry relations, which are replaced by a more complicated
non-linear differential equation.

In the magnetic string case our asymptotically flat solutions
extend the near horizon attractor geometries we found in
\CastroSD. As described in \CastroSD, from the near horizon
AdS$_3$ solution we can read off the central charges of the dual
CFT, and thereby confirm an earlier result based on anomalies and
supersymmetry \refs{\us,\KrausZM}.  In the full solution the
metric and matter fields exhibit oscillatory behavior of the same
sort as found in \refs{\DabholkarDQ,\senstretch}.    As mentioned
above, the simplest example is the single charge solution
corresponding to a heterotic string with vanishing momentum (or,
in the language of M-theory on $K3\times T^2$, an M5-brane wrapped
on $K3$). This provides an example of how $R^2$ corrections can
replace a naked singularity with a smooth, zero entropy, geometry.

In the electric case, after constructing the black hole solutions
we discuss their entropy.  Given the near horizon AdS$_2$
geometry, the computation of the Bekenstein-Hawking-Wald entropy
reduces to evaluating a particular function at its extremum. The
resulting entropy formula turns out to be extremely simple:
expressed in terms of the charges, it takes the same form as in
the two derivative case except that the charges are shifted by an
amount proportional to the second Chern class of the Calabi-Yau.
With a natural definition of horizon area, the formula $S=A/4$
continues to hold.

We compare our corrected entropy formula to a previously
conjectured result \GuicaIG\ based on the 4D-5D connection and the
topological string free energy.  The results precisely agree when
expressed in terms of the electric potentials.  Expressed instead
in terms of the electric charges, there is a mismatch.  Our
entropy formula also agrees quantitatively with a correction found
by Vafa in terms of a microscopic model  of 5D black holes in
M-theory on an elliptically fibred Calabi-Yau \VafaGR.

The remainder of this paper is organized as follows. In section 2
we review the construction of higher derivative terms using the
off-shell formalism. In section 3 we find supersymmetric solutions
of magnetic type, interpreted as string solutions. We establish
that 5D strings interpolate smoothly between an AdS$_3\times S^2$
near horizon geometry and an asymptotically flat region. In
section 4 we find supersymmetric solutions of electric type. We
discuss their near horizon behavior and we establish that they
interpolate smoothly between an AdS$_2\times  S^3$ near horizon
geometry and an asymptotically flat region. In section 5 we
discuss the entropy of the 5D Calabi Yau black holes.

\newsec{5D supergravity with $R^2$ corrections}

We begin with a brief review of higher-derivative corrections to
$N =2$ supergravity in five dimensions \HanakiPJ. We use the
superconformal formalism,  developed in
\refs{\KugoHN,\BergshoeffHC}, which can be gauge-fixed to the
familiar Poincare supergravity.  Our conventions are summarized in
appendix A.

\subsec{The supersymmetry transformations}

Before introducing the specific action that we analyze in this
paper, let us briefly discuss the relevant supersymmetry
multiplets. The irreducible  Weyl multiplet contains the fields:
\eqn\aaa{ e_\mu^{~a}~, ~~~\psi_\mu~, ~~~V_\mu~, ~~~b_\mu~, ~~~v^{ab}, ~~~\chi~, ~~~D~.
}
The first two fields are the vielbein and gravitino. The  $V_\mu$
is the vector boson associated with the gauging of the $SU(2)$
R-symmetry under which all fermionic variables and fields
transform\foot{We suppress the $SU(2)$ indices of $V_\mu$ and all
fermionic variables because they play no role in our work.}, while
$b_\mu$ is the gauge field of dilatational symmetry. We will
ignore these gauge fields in the future, for they are gauged way
when one reduces to Poincare supergravity. Finally are three
auxiliary fields: an anti-symmetric tensor $v_{ab}$,
the fermion $\chi$, and the scalar $D$. The vector multiplet
consists of the gauge field $A_\mu^I$, the scalar $M^I$, the
gaugino $\Omega^I$, and also a scalar $Y^{I}$, which will be
gauged away.

Also of importance is the hypermultiplet. Although the hypers
decouple from the physics we are developing, by gauge-fixing them
we effectively  couple the supersymmetry variations of the
irreducible Weyl and vector multiplets \CastroSD. We will ignore
the hypermultiplets henceforth and take the coupled supersymmetry
variations as a starting point.

Since the Weyl and vector multiplets are irreducible representations, the
variations of the fields under supersymmetry transformations are independent
of the action of the theory under consideration. As usual, a bosonic field configuration is
supersymmetric when all fermion variations vanish. The supersymmetry conditions
from the fermion variations are
\eqn\ab{\eqalign{
\delta\psi_\mu&=\left({\cal D}_\mu+{1\over2}v^{ab}\gamma_{\mu ab}-{1\over3}\gamma_\mu\gamma\cdot
v\right)\epsilon=0~, \cr
\delta\Omega^{I}&=\left(-{1\over4}\gamma\cdot
F^I-{1\over2}\gamma^a\partial_aM^I-{1\over3}M^I\gamma\cdot v
\right)\epsilon=0~,\cr
\delta \chi &=\left(D-2\gamma^c\gamma^{ab}{\cal
D}_av_{bc}-2\gamma^a\epsilon_{abcde}v^{bc}v^{de}+
{4\over3}(\gamma\cdot v)^2\right)\epsilon=0~,
}}
where $\gamma \cdot T=\gamma_{ab} T^{ab}$ for a rank-2 tensor $T_{ab}$.

Once again, we wish to emphasize that the above variations are independent of the action of the theory. Indeed, this is the whole point of retaining the auxiliary fields. Consequently, the equations \ab\ serve as the supersymmetry conditions in the presence of higher-derivative terms.

\subsec{The two-derivative action}

After gauge fixing to Poincare supergravity the two-derivative Lagrangian constructed
from the Weyl multiplet and $n_V$ vector multiplets reads
\eqn\ac{\eqalign{{\cal L}_0=&
-{1\over2}D-{3\over4}R+v^2+{\cal
N}\left({1\over2}D-{1\over4}R+3v^2\right)+2{\cal
N}_Iv^{ab}F^I_{ab}\cr &+{\cal
N}_{IJ}\left({1\over4}F^I_{ab}F^{Jab}+{1\over2}\partial_aM^I\partial^aM^J\right)+{1\over24e}
c_{IJK}A^I_{a}F^{J}_{bc}F^{K}_{de}\epsilon^{abcde}~.
}}
One can integrate out the auxiliary fields $D$ and $v_{ab}$ by
solving their equations of motion and substituting the solutions
back into \ac. This yields the familiar $N=2$ Lagrangian arising
from the compactification of eleven-dimensional supergravity on a
Calabi-Yau manifold with intersection numbers $c_{IJK}$
\CadavidBK.

The functions defining the scalar manifold are
\eqn\aab{{\cal
N}={1\over6}c_{IJK}M^IM^JM^K~,\quad {\cal N}_I=\partial_I{\cal
N}={1\over2}c_{IJK}M^JM^K~,\quad {\cal N}_{IJ}=c_{IJK}M^K~,}
where $I,J,K= 1, \ldots, n_V$. At the two-derivative level, the $D$ equation of motion
imposes the constraint ${\cal N}=1$ defining real special geometry.
However, higher derivative corrections make the geometry
of the scalar moduli space more complicated.

\lref\DuffWD{
  M.~J.~Duff, J.~T.~Liu and R.~Minasian,
  ``Eleven-dimensional origin of string / string duality: A one-loop test,''
  Nucl.\ Phys.\  B {\bf 452}, 261 (1995)
  [arXiv:hep-th/9506126].
}

\subsec{The four-derivative action}

A particular four-derivative term is special in that its coefficient is
determined by M5-brane anomaly cancellation via
anomaly inflow \DuffWD.  This is the mixed gauge-gravitational
Chern Simons term
\eqn\aha{ e{\cal L}_{\rm CS} = {c_{2I}\over 24\cdot 16}
\epsilon_{abcde} A^{Ia} R^{bcfg}R^{de}_{~~fg}~.}
It is believed that all four derivative terms are related to this term by
supersymmetry \HanakiPJ.
The supersymmetric completion of the term \aha\ was derived in \HanakiPJ.
The bosonic terms are
\eqn\af{\eqalign{{\cal L}_1& = {c_{2I} \over 24} \Big( {1\over
16e} \epsilon_{abcde} A^{Ia} C^{bcfg}C^{de}_{~~fg} + {1 \over
8}M^I C^{abcd}C_{abcd} +{1 \over 12}M^I D^2 +{1 \over
6}F^{Iab}v_{ab}D \cr &+{1 \over 3}M^I C_{abcd}
v^{ab}v^{cd}+{1\over 2}F^{Iab} C_{abcd} v^{cd} +{8\over 3}M^I
v_{ab} \hat{\cal D}^b \hat{\cal D}_c v^{ac} \cr &+{4\over 3} M^I
{\hat{\cal D}}^a v^{bc} {\hat{\cal D}}_a v_{bc} + {4\over 3} M^I
{\hat{\cal D}}^a v^{bc} {\hat{\cal D}}_b v_{ca} -{2\over 3e} M^I
\epsilon_{abcde}v^{ab}v^{cd}{\hat{\cal D}}_f v^{ef}\cr &+ {2\over
3e} F^{Iab}\epsilon_{abcde}v^{cf} {\hat{\cal D}}_f v^{de}
 +e^{-1}F^{Iab}\epsilon_{abcde}v^c_{~f}{\hat{\cal D}}^d v^{ef}\cr
& -{4 \over 3}F^{Iab}v_{ac}v^{cd}v_{db}-{1 \over 3}
F^{Iab} v_{ab}v^2 +4 M^I v_{ab} v^{bc}v_{cd}v^{da}-M^I
(v^2)^2\Big)~.}}
Here $C_{abcd}$ is the Weyl tensor. The superconformal derivative
is related to the usual derivative as ${\hat{\cal D}}_\mu = {\cal
D}_\mu  - b_\mu$. In our gauge the dilatational connection $b_\mu$
vanishes. However, its derivative does not vanish so the second
superconformal covariant derivative is nontrivial, {\it viz.}
\eqn\ah{v_{ab}\hat{{\cal D}}^b\hat{{\cal D}}_cv^{ac}=v_{ab}{\cal
D}^b{\cal
D}_cv^{ac}-{2\over3}v^{ac}v_{cb}R_{a}^{\phantom{a}b}-{1\over12}v_{ab}v^{ab}R~.}

The complete action \af\ is evidently somewhat unwieldy.
Fortunately, most terms play no role
 in our applications. For example, the parity-odd terms (proportional to
$\epsilon_{abcde}$) vanish on our spacetimes. Their significance
is that they fix the normalization through the anomaly term \aha.
We will ultimately need just the equations of motion for
$D$ and for the gauge fields, and these involve just a few of the
terms in \af.

\newsec{Magnetic solutions:  strings
with AdS$_3 \times S^2$ near horizon geometry}

Our strategy for finding regular solutions in the higher
derivative theory is to first write an {\it ansatz} consistent
with the assumed symmetries, and then demand unbroken
supersymmetry.  This part of the analysis proceeds the same
whether we consider the two derivative or four derivative theory,
and hence is quite manageable.  Supersymmetry does not completely
determine the solution, however --- we also need to impose the
Bianchi identity and the special geometry constraint, the latter
coming from the $D$ equation of motion.  Only at this last stage
do we need specific information about the action.

\subsec{Ansatz: magnetic background}

We are interested in higher-derivative corrections to the supersymmetric black string solutions
carrying magnetic charges $p^I$ studied in \ChamseddineQS. We assume translation invariance
along the string, and spherical symmetry in the transverse
directions.  To make these symmetries explicit, we write our {\it
ansatz} as
\eqn\ba{ds^2=e^{2U_1(r)}\left(dt^2-dx_4^2\right)-e^{-4U_2(r)}dx^i
dx^i~,\quad dx^i dx^i =dr^2+r^2d\Omega_2^2~,}
where $i=1,2,3$. The two-forms $F^I$ and $v$ will  be proportional
to the volume form on $S^2$. We chose the vielbein as
\eqn\bab{\eqalign{e^{\hat{a}}=&e^{U_1}dx^a~,~~\quad a=0,4~,\cr
e^{\hat{i}}=&e^{-2U_2}dx^i~,\quad i=1,2,3~.}}
The non-trivial spin connections are
\eqn\bac{
\omega_a^{~\hat{a}\hat{i}} = - e^{U_1+2U_2} \partial_i U_1~, ~~~~~~\omega_k^{~\hat{i}\hat{j}} = 2 \delta_k^i \partial_j U_2 - 2 \delta_k^j \partial_i U_2~.}

\subsec{Supersymmetry conditions}
We start by analyzing the supersymmetry conditions \ab\ in the
background \ba.   The supersymmetry parameter $\epsilon$ is
constant along the string and obeys
\eqn\bb{\gamma_{\hat{t}\hat{4}}\epsilon=-\epsilon~.}

\vskip.2cm \noindent {\bf Gravitino variation} \vskip .2cm

We first analyze the gravitino variation \ab\
\eqn\bc{\delta\psi_\mu=\left({\cal D}_\mu+{1\over2}v^{ab}\gamma_{\mu
ab}-{1\over3}\gamma_\mu\gamma\cdot v\right)\epsilon=0~.}
For the background \ba, the covariant derivative is
\eqn\bca{\eqalign{{\cal
D}_a&=\partial_a-{1\over2}e^{U_1+2U_2}\partial_iU_1\gamma_{\hat{a}\hat{i}}~,\cr{\cal
D}_i&=\partial_i+\partial_jU_2\gamma_{\hat{i}\hat{j}}~.}}
Along the string, equation \bc\ simplifies to
\eqn\bcb{\left[-{1\over2}e^{U_1+2U_2}\partial_iU_1\gamma_{\hat{a}\hat{i}}+
{1\over6}e^{U_1}v_{\hat{i}\hat{j}}\gamma_{\hat{a}\hat{i}\hat{j}}\right]\epsilon=0~.}
It is convenient to use the projection \bb\ in the form
\eqn\bcd{\gamma_{\hat{i}\hat{j}\hat{k}} \epsilon = -\varepsilon_{ijk} \epsilon~,}
where $\varepsilon_{123}=1$. Then
\eqn\bce{
\gamma_{\hat{i}\hat{j}} \epsilon = \gamma^{\hat{k}} \gamma_{\hat{i}\hat{j}\hat{k}} \epsilon = \varepsilon_{ijk}\gamma_{\hat{k}} \epsilon~.}
So \bcb\ becomes
\eqn\bcf{
\left[-{1\over2} e^{U_1+2U_2} \partial_k U_1 + {1\over6} e^{U_1} v_{\hat{i}\hat{j}} \varepsilon_{ijk} \right] \gamma_{\hat{a}\hat{k}} \epsilon=0~,}
from which we can solve for the auxiliary field,
\eqn\bcg{
v_{\hat{i}\hat{j}} = {3\over2} e^{2U_2} \varepsilon_{ijk} \partial_k U_1~,}
or in coordinate frame
\eqn\bch{
v_{ij} = {3\over2} e^{-2U_2} \varepsilon_{ijk} \partial_k U_1~.}

Consider now the components of the gravitino variation along
$x^i$,
\eqn\bci{
\left[\partial_i+\partial_jU_2\gamma_{\hat{i}\hat{j}} + {1\over2} v_{\hat{j}\hat{k}}\left( \gamma_{i\hat{j}\hat{k}} - {2\over3} \gamma_i \gamma_{\hat{j}\hat{k}} \right)\right]\epsilon =0~.
}
The $v_{\hat{j}\hat{k}}$ terms split into a ``radial'' part where either $j,k$ is equal to $i$, and an ``angular'' part where $i\ne j \ne k$. Thus we have two conditions
\eqn\bcj{\eqalign{
0 &= \left(\partial_i - {1\over6} \varepsilon_{ijk} e_i^{~\hat{i}} v_{\hat{j}\hat{k}} \right) \epsilon~,  \cr
0 &= \left( \partial_j U_2 \varepsilon_{ijk} + {2\over3} v_{\hat{i}\hat{k}} e_i^{\hat{i}}\right)\gamma_{\hat{k}} \epsilon ~,
}}
where there is no summation over $i$. The second equation leads to $U_2 = U_1$, so we will drop the subscripts on $U$ from now on. The first equation then leads to
\eqn\bck{
\left(\partial_i -\half \partial_i U \right) \epsilon =0~,
}
and so the Killing spinor takes the form
\eqn\bcl{
\epsilon = e^{U/2} \epsilon_0~,
}
where $\epsilon_0$ is some constant spinor.

It will be convenient  to use cylindrical coordinates from now on.
The metric takes the form
\eqn\bcm{ds^2=e^{2U}\left(dt^2-dx_4^2\right)-e^{-4U}(dr^2+r^2d\Omega_2^2)~,}
in terms of a single function $U(r)$. The coordinate frame expression \bch\ is a tensor statement on the 3-dimensional base space, where $\varepsilon_{abc}$ is a completely anti-symmetric tensor with components $\pm\sqrt{g}$. So in cylindrical coordinates the
auxiliary two-form is
\eqn\bcn{v_{\theta\phi}={3\over2}e^{-2U}r^2\sin\theta\partial_rU~,\quad v_{\thetah\phih}={3\over2}e^{2U}\partial_rU~,}
with other components vanishing due to spherical symmetry in the transverse space. The projection \bb\ in cylindrical coordinates can be written as
\eqn\bco{\gamma_{\rh\thetah\phih} \epsilon =- \epsilon~.}

\vskip.2cm \noindent {\bf Gaugino variation} \vskip .2cm

Evaluated on the magnetic background, the gaugino
variation $\delta\Omega^I$  in \ab\ gives
\eqn\bd{\left(\gamma_{\hat{\theta}\hat{\phi}}
F^{I\hat{\theta}\hat{\phi}}+\gamma^{\hat{r}}e^r_{\hat{r}}\partial_rM^I+{4\over3}M^I\gamma_{\hat{\theta}\hat{\phi}}
v^{\hat{\theta}\hat{\phi}}\right)\epsilon=0~.}
Using \bco\ and solving for the field strength we get
\eqn\bda{\eqalign{F^{I\hat{\theta}\hat{\phi}}&=e^{2U}\partial_rM^I-{4\over3}M^Iv^{\hat{\theta}\hat{\phi}}\cr
&=\partial_r(M^Ie^{-2U})e^{4U}~.}}
In coordinate frame,
\bda\ becomes
\eqn\bde{F^I_{\theta\phi}=\partial_r(M^Ie^{-2U})r^2\sin\theta~.}
This equation is the first hint of the expected attractor behavior: the flow of
the scalars $M^I$ is completely determined by the magnetic field $F^I$.

\vskip.2cm \noindent {\bf Auxiliary fermion variation} \vskip .2cm

The last supersymmetry variation
to solve is $\delta\chi=0$. Neglecting the $\epsilon$-terms since we look for parity invariant solutions, this
condition is
\eqn\be{\left(D-2\gamma^c\gamma^{ab}{\cal
D}_av_{bc}+{4\over3}(\gamma\cdot v)^2\right)\epsilon=0~.}
The relevant components of the covariant derivative of $v$ for the
contraction in \be\ are
\eqn\beaa{{\cal D}_\theta v_{r\phi}={\cal D}_\phi
v_{\theta r}=-\Gamma^{\theta}_{\theta r}v_{\theta\phi}~,~ {\cal D}_r
v_{\theta\phi}=\partial_r v_{\theta\phi}-2\Gamma^{\theta}_{\theta
r}v_{\theta\phi}~,}
with
\eqn\beab{\Gamma^{\theta}_{\theta r}=\Gamma^{\phi}_{\phi
r}=-2\partial_rU+{1\over r}~.}
Then, the second term in \be\ becomes
\eqn\bea{\eqalign{\gamma^c\gamma^{ab}{\cal
D}_av_{bc}&=e^r_{\hat{r}}e^\theta_{\hat{\theta}}e^\phi_{\hat{\phi}}\left(-4{\cal
D}_\theta v_{r\phi}+2{\cal
D}_rv_{\theta\phi}\right)\gamma^{\hat{r}\hat{\theta}\hat{\phi}}\cr&=2{e^{6U}\over
r^2\sin\theta}\partial_rv_{\theta\phi}\gamma^{\hat{r}\hat{\theta}\hat{\phi}}
~, \cr
&={3\over2}e^{6U}\nabla^2(e^{-2U})\gamma_{\hat{r}\hat{\theta}\hat{\phi}}~,}}
with
$\nabla^2=\partial_i\partial_i=r^{-2}\partial_r(r^2\partial_r)$ due to spherical symmetry.
Inserting \bea\ in \be\ we have
\eqn\beb{\left(D-3e^{6U}\nabla^2(e^{-2U})\gamma_{\hat{r}\hat{\theta}\hat{\phi}}
-{16\over3}(v_{\hat{\theta}\hat{\phi}})^2\right)\epsilon=0~,}
where we used
\eqn\beba{(\gamma\cdot v)^2=-4(v_{\hat{\theta}\hat{\phi}})^2~.}
Using the projection \bco\ and substituting the auxiliary field \bcn\ into \beb\ we find
\eqn\bec{\eqalign{D&=-3e^{6U}\nabla^2(e^{-2U})+{16\over3}(v_{\hat{\theta}\hat{\phi}})^2\cr
&=3e^{6U}\left(-\nabla^2(e^{-2U})+4e^{-2U}(\nabla
U)^2\right)\cr&= 6 e^{4U}\nabla^2U~.}}

What we have found so far is that supersymmetry demands a metric
of the form \bcm, an auxiliary two tensor of the form \bcn, the
gauge field strengths \bde, and the auxiliary D-field \bec. All
told the entire solution is now specified in terms of the
functions $M^I$ and $U$ which are not fixed by supersymmetry
alone.

\subsec{Equations of motion}

Having exhausted the implications of unbroken supersymmetry, we
now need to use information from the equations of motion.

\vskip.2cm \noindent {\bf Maxwell's equations} \vskip .2cm

Any specific string solution is parameterized by the  values of the
magnetic charges as measured by surface integrals at infinity.
These in turn determine the gauge fields in the interior via the
Maxwell equations.

We first consider the equation of motion
\eqn\bha{\partial_{\theta}\left(\sqrt{g}{\partial {\cal
L}\over\partial F^I_{\theta\phi}}\right)=0~.}
Spherical symmetry implies that the expression in parenthesis is a
function of $r$ only, hence \bha\ is satisfied identically for any
field strength  $F^I_{\theta\phi}={\cal F}^I(r)\sin\theta$. Thus
we get no new information from this equation of motion.

In the magnetic case the nontrivial condition arises from the Bianchi identity $dF^I=0$.
The point is that the expression \bda\ for $F^I$ determined from supersymmetry
is not automatically a closed form. Therefore, the Bianchi identity
\eqn\bia{
\partial_rF^I_{\theta\phi}=
\partial_r\left(r^2\partial_r(M^Ie^{-2U})\right)\sin\theta=
0~,}
is nontrivial. Physically, this is because supersymmetry is
consistent with any extended distribution of magnetic charges,
while here we are demanding the absence of charge away from the
origin.  The equation \bia\ integrates to
\eqn\bic{r^2\partial_r(M^Ie^{-2U})=-{p^I\over 2}~,}
where $p^I$ is the quantized magnetic charge carried by $F^I$. We note that
the field strength
\eqn\bid{F^I=-{p^I\over2}\epsilon_2~,}
does not get modified after including higher derivatives since it
is topological.

The solutions to \bic\ are harmonic functions on the
three-dimensional base space. We are just interested in the
simplest solution
\eqn\bie{M^Ie^{-2U}=H^I=M^I_\infty+{p^{I}\over 2r}~,}
with $M^I_\infty$ the value of $M^I$ in the asymptotically flat region
where $U=0$.

\vskip.2cm \noindent {\bf D equation} \vskip .2cm

So far, by imposing the conditions for supersymmetry and integrating
the Bianchi identity, we have been able to write our solution in terms
of one unknown function $U(r)$. To determine this remaining
function we use the equation of motion for the auxiliary field $D$.
Inspecting \ac\ and \af\  we see that the only D-dependent terms
in the Lagrangian are
\eqn\bl{{\cal L}_D={1\over2}({\cal
N}-1)D+{c_{2I}\over24}\left({1\over12}M^ID^2+{1\over6}F^{Iab}v_{ab}D\right)~.}
Therefore, the equation of motion for $D$ is
\eqn\bk{{\cal
N}=1-{c_{2I}\over72}\left(F^I_{ab}v^{ab}+M^ID\right)~.}
Inserting the gauge-field \bde, the auxiliary field \bcn, and the
D-field \bec\ gives
\eqn\bka{ e^{-6U}={1\over
6}c_{IJK}H^IH^JH^K+{c_{2I}\over24}\left(\nabla H^I\nabla
U+2H^I\nabla^2U\right)~.}
Here $H^I$ are the harmonic functions defined in \bie\ and we used
\eqn\bkb{{\cal N}={1\over 6}c_{IJK}H^IH^JH^Ke^{6U}~.}
The D  constraint \bka\ is now an ordinary differential equation
that determines $U(r)$. Its solution specifies the entire geometry
and  all the matter fields.

We can solve \bka\ exactly in the near horizon region. This case
corresponds to vanishing integration constants in \bie\ so that
\eqn\bmd{H^I = {p^I\over 2r}~.}
Then \bka\ gives
\eqn\bmdb{e^{-6U}={1\over8r^3}\left(p^3+{1\over12}c_{2}\cdot
p\right)={\ell_S^3\over r^3}~,}
where $p^3 = {1\over 6} c_{IJK} p^I p^J p^K$. The geometry in this
case is AdS$_3\times S^2$ with the scale $\ell_S$ in agreement
with our previous work \CastroSD.

The asymptotically flat solutions to \bka\ cannot in general be
found in closed  form. In the following two subsections we discuss
an approximate solution and an example of numerical integration.

\subsec{Corrected geometry for large black strings} One way to
find solutions to \bka\ is by perturbation theory. This strategy
captures the correct  physics when the solution is regular already
in the leading order theory, {\it i.e.} for large black strings.
Accordingly, the starting point is the familiar solution
\eqn\bm{e^{-6U_0}={1\over6}c_{IJK}H^IH^JH^K~,}
to the two-derivative theory. This solves \bka\ with $c_{2I}=0$.

Although $c_{2I}$ is not small it will be multiplied by terms that
are of higher order in the derivative expansion. It is therefore
meaningful to expand the full solution to \bka\ in the form
\eqn\bme{e^{-6U}=e^{-6U_0}+c_{2I}\varepsilon^I+
{1\over2}c_{2I}c_{2J}\varepsilon^{IJ}+\ldots~,}
where $\varepsilon^I(r), \varepsilon^{IJ}(r),\ldots$ determine the
corrected geometry with increasing precision.

Inserting \bme\ in \bka\ and keeping only the terms linear in
$c_{2I}$ we find the first order correction\foot{It is understood
that the correction $\varepsilon^I$ is only defined in the
combination $c_{2I}\varepsilon^I$.}
\eqn\bmb{\varepsilon^I={1\over24}(\nabla H^I\nabla
U_0+2H^I\nabla^2U_0)~.}
Iterating, we find the second order correction
\eqn\bmg{\varepsilon^{IJ}=-{1\over72}\left(\nabla
H^I\nabla(e^{6U_0}\varepsilon^J)+2H^I\nabla^2(e^{6U_0}\varepsilon^J)
\right)~,}
where the first order correction $\varepsilon^I$ is given by \bmb.
Higher orders can be computed similarly. In summary, we find that
starting from a smooth solution to the two-derivative theory we
can systematically and explicitly compute the higher order
corrections. The series is expected to be uniformly convergent.

In the near horizon limit \bmd\ the full solution \bmdb\ is
recovered exactly when taking the leading correction \bmb\ into
account. As indicated in \bmdb\ the effect of the higher
derivative corrections is to expand the sphere by a specific
amount (which is small for large charges). The perturbative
solution gives approximate expressions for the corrections also in
the bulk of the solution. Numerical analysis indicates that the
corrections remain positive so at any value of the isotropic
coordinate $r$ the corresponding sphere is expanded by a specific
amount.

\subsec{Fundamental strings}

\lref\SenCJ{
  A.~Sen,
  ``String String Duality Conjecture In Six-Dimensions And Charged Solitonic
  Nucl.\ Phys.\  B {\bf 450}, 103 (1995)
  [arXiv:hep-th/9504027].
}

\lref\HarveyRN{
  J.~A.~Harvey and A.~Strominger,
  ``The Heterotic String Is A Soliton,''
  Nucl.\ Phys.\  B {\bf 449}, 535 (1995)
  [Erratum-ibid.\  B {\bf 458}, 456 (1996)]
  [arXiv:hep-th/9504047].
} One of the main motivations for developing higher derivative
corrections is their potential to regularize geometries that are
singular in the lowest order supergravity approximation
\refs{\DabholkarYR,\DabholkarDQ,\senstretch\DDMP,\us}. This is the
situation for small strings, by which we mean charge
configurations satisfying $p^3 = {1\over 6}c_{IJK}p^I p^J p^K=0$.

A particularly important example of a small string is when the
Calabi-Yau is $K3\times T^2$ and the only magnetic charge that is
turned on is the one corresponding to an $M5$-brane wrapping the
$K3$. The resulting 5D string is then dual, via IIA-heterotic
duality,  to the fundamental heterotic string
\refs{\SenCJ,\HarveyRN}.

Let $M^1$ be the single modulus on the torus and $M^i$ be the
moduli of $K3$ where $i=2,\ldots,23$. The charge configuration of
interest specifies the harmonic functions as
\eqn\bn{\eqalign{ H^1&=M^1_{\infty}+{p^1\over2r}~,\cr
H^{i}&=M^i_\infty~,~~~~~~i=2,\ldots,23~. }}
The only nonvanishing intersection numbers are $c_{1ij}=c_{ij}$
where $c_{ij}$ is the intersection matrix for $K3$.   We choose
$M^i_\infty$ consistent with
\eqn\bna{{\cal N}e^{-6U}={1\over6}c_{IJK}H^IH^JH^K=H^1~.}
The master equation \bka\ now becomes
\eqn\bnb{H^1 = e^{-6U}- \left[ \partial_r H^1\partial_r
U+2H^1~{1\over r^2}\partial_r (r^2 \partial_r U)\right] ~,}
where we used $c_2(K3)=24$ and $c_{2i}=0$. We can write this more
explicitly as
\eqn\bnc{ 1 + {p^1\over 2r} = e^{-6U}  - 2(1 + {p^1\over 2r})
U^{\prime\prime} -{4\over r}\left( 1 + {3p^1\over 8r} \right)
U^\prime~, }
where primes denote derivatives with respect to $r$.

In our units distance $r$ is measured in units of the 5D Planck length.
The parameter $p^1$ is a pure number counting the fundamental strings.
We take $p^1\gg 1$ so as to have an expansion parameter.
We will analyze the problem one region at a time.

\vskip.2cm \noindent {\bf The AdS$_3\times S^2$-region} \vskip
.2cm

This is the leading order behavior close to the string. According
to our near horizon solution \bmdb\ we expect the precise
asymptotics
\eqn\bnd{ e^{-6U} \to {\ell^3_S\over r^3}~~~,~r\to 0~,
}
where the $S^2$-radius is given by \eqn\bne{ \ell_S = \left(
{p^1\over 4}\right)^{1/3}~. } Since we assume $p^1\gg 1$ this is
still much larger than the 5D Planck scale. The modulus describing the
volume of the internal $T^2$ is
\eqn\bnea{ M^1 = {p^1\over
2\ell_S} = 2^{-1/3}(p^1)^{2/3}~, }
which also corresponds to the length scale $(p^1)^{1/3}$.

\vskip.2cm \noindent {\bf The near-string region} \vskip .2cm

We next seek a solution in the entire range $r\ll p^1$ which
includes the scale \bne\ but reaches further out. In fact, it may
be taken to be all of space in a scaling limit where
$p^1\to\infty$.

In the near string region \bnc\ reduces to
\eqn\bnf{ {p^1\over 2r}
= e^{-6U}  - {p^1\over r} U^{\prime\prime} -  {3p^1\over 2r^2}
U^\prime~. }
We can scale out the string number $p^1$ by
substituting
\eqn\bng{ e^{-6U(r)} = {p^1\over 4r^3}
e^{-6\Delta(r)}~, }
which amounts to
\eqn\bnh{ U(r) = {1\over 2}
\ln {r\over\ell_S} + \Delta(r)~. }
This gives
\eqn\bni{
\Delta^{\prime\prime} + {3\over 2r} \Delta^\prime   + {1\over
4r^2} ( 1 - e^{-6\Delta}) + {1\over 2}=0~, }
which describes the
geometry in the entire region $r\ll p^1$. The asymptotic behavior
at small $r$ is
\eqn\bnj{ \Delta(r) = -{1\over 13}r^2 + {3\over (13)^3}r^4 +\cdots~.
}
Since $\Delta(r)\to 0$ smoothly as $r\to 0$
we have an analytical description of the approach to the
AdS$_3\times S^2$ region.

The asymptotic behavior for large $r$ is also smooth. Expanding in
$u={1\over r}$ we find
\eqn\bnk{ \Delta(r) = -{1\over 6} \ln (2r^2) -  {1\over
36r^2}+\cdots~. }
It is straightforward to
solve \bni\ numerically. Figure 1 shows the curve that interpolates
between the asymptotic forms \bnj\ and \bnk. The oscillatory behavior
in the intermediate region is characteristic of higher derivative theories.
We comment in more detail below.

\fig{Analytical and numerical results for $\Delta(r)$ in the near
string region, $r\ll p^1$. In both plots the blue curve is given
by solving \bni\ numerically. {\it Left}: the numerical solution
close to the string with the approximate solution \bnj\ given in
red. {\it Right}: the numerical solution further away with the
approximate solution \bnk\ given in green. The plots have
overlapping values of $r$ but different
scales.}{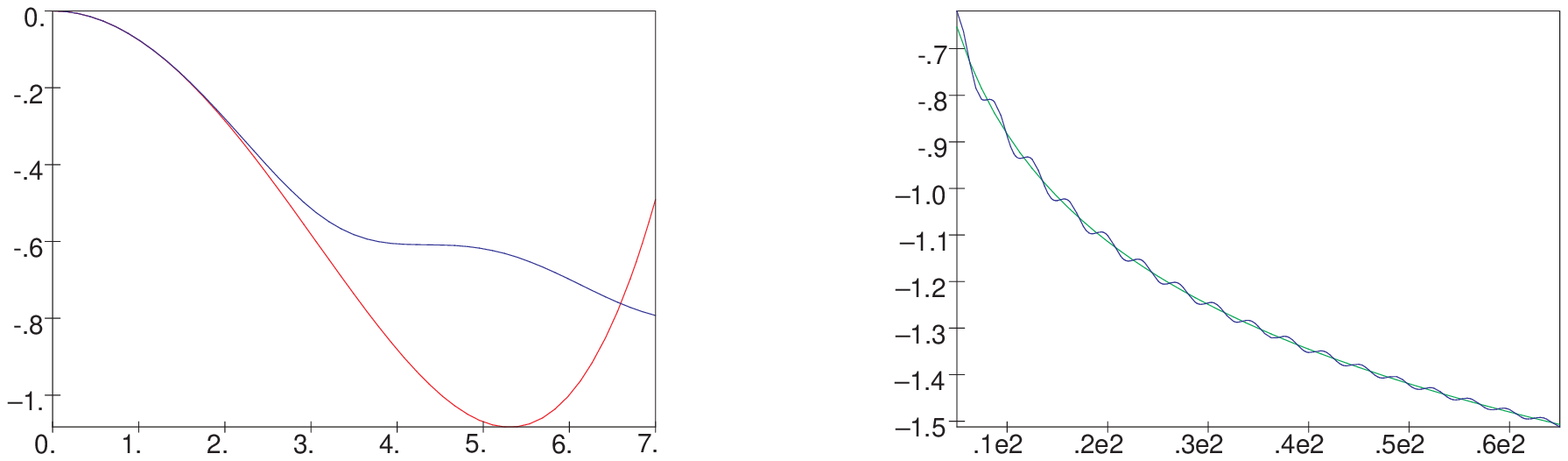}{6.5truein}

In the original variable $U(r)$ the approximation \bnk\ gives
\eqn\bnl{ e^{-6U} = {p^1\over 2r} \left( 1 + {1\over
6r^2}+\ldots\right)~, }
for large $r$. The leading behavior agrees
with the near string behavior $e^{-6U} = H^1\sim {p^1\over 2r}$
familiar from the description of a fundamental string in
two-derivative supergravity. In the full theory this singular
region is replaced by a smooth geometry described by $\Delta(r)$.

\vskip.2cm \noindent {\bf The approach to asymptotically flat
space} \vskip .2cm

We still need to analyze the region where $r$ is large, meaning
$r\sim p^1$ or larger. Although standard two-derivative
supergravity is expected to describe this region it is instructive
to consider the possible corrections.

In the asymptotic region the full equation \bnb\ simplifies to
\eqn\bnm{ 1 + {p^1\over 2r} = e^{-6U} - 2( 1+ {p^1\over 2r})
U^{\prime\prime}~. }
Terms with explicit factors of $1/r$ were
neglected but we kept derivatives with respect to $r$, to allow
for structure on Planck scale even though $r\sim p^1\gg 1$.
Changing variables as
\eqn\bnn{ e^{-6U} = (1 + {p^1\over 2r})
e^{-6W}~, } we find \eqn\bno{ W^{\prime\prime}= {1\over 2}(
e^{-6W} -1) \simeq -3W~. }
The expansion for small $W$ is justified because
\bnl\ imposes the boundary condition $W\to 0$ for $r\ll p^1$.

The solution $W=0$ expected from supergravity is in fact a
solution to \bno\ but there are also more general solutions of the
form
\eqn\bnp{ W = A \sin ( \sqrt{3} r +\delta)~. }
The amplitude of this solution is undamped, so it is not really an
intrinsic feature of the localized string solution we consider.
Instead it is a property of fluctuations about flat space, albeit an
unphysical one. The existence of such spurious solutions is a
well-known feature of theories with higher derivatives, and is
related to the possibility  of field redefinitions
\refs{\ZwiebachUQ, \DabholkarDQ, \senstretch}. In the present
context the issue is that other variables such as $\tilde{W} =
(\nabla^2 - 3)W$ exhibit no spurious solutions. It would be
interesting to make this interpretation of the spurious solutions
more explicit. It would also be interesting to understand possible
relations of our fundamental string solutions with the picture
proposed in \GiveonPR.

\newsec{Electric solutions:  black holes with AdS$_2\times S^3$ near horizon geometry}

We now consider the case of electrically charged,  spherically
symmetric solutions. We follow the same strategy as in the
analysis of the magnetic solutions: we start from an {\it ansatz}
with the desired symmetry, then use the supersymmetry conditions
\ab\ to relate various functions in the {\it ansatz}, and finally
impose appropriate equations of motion to obtain the full
solution. The solutions we study are the higher-derivative corrected versions of those in \SabraYD.

In the electric case  we start with a metric of the form
\eqn\ca{ ds^2 = e^{4U_1(x)} dt^2 - e^{-2U_2(x)} dx^i dx^i~, }
where $i=1\ldots4$. This gives the vielbein
\eqn\cab{
e^{\th} = e^{2U_1} dt~,~~~~~e^{\hat{i}} = e^{-U_2} dx^i~,
}
and spin connections \eqn\cac{ \omega_t^{~\hat{t}\hat{i}}=-2
\partial_i U_1 e^{2U_1+U_2}~, ~~~~~\omega_k^{~\hat{i}\hat{j}}
=\partial_j U_2 \delta_k^{~i}-\partial_i U_2 \delta_k^{~j}~. } In
this paper we limit ourselves to spherically  symmetric solutions
for which
\eqn\cad{
v^{\hat{i}\hat{j}}=0~.}

\subsec{Supersymmetry conditions} We now make the conditions \ab\
imposed by  supersymmetry explicit for our electric {\it ansatz}.

\vskip.2cm \noindent {\bf Gravitino variation} \vskip .2cm

We begin with  constraints from the gravitino variation, the first
equation in \ab. The temporal part of the gravitino variation
reads
\eqn\cbaa{
\left[\partial_t+{1\over 2}\omega_t^{~\hat{t}\hat{i}}\gamma_{\hat{t}\hat{i}} + {1\over 2} v^{ab}\left(\gamma_{t ab}-{2\over3}\gamma_t\gamma_{ab} \right) \right]\epsilon=0~.
}
We assume that the Killing spinor is time-independent and
satisfies the projection
\eqn\cba{
\gamma_{\hat{t}} \epsilon = -\epsilon~.
}
Inserting \cad\ we find
\eqn\cbc{
\left( {1\over 2} \omega_t^{~\hat{t}\hat{i}} -{2\over3} e_t^{~\hat{t}} v^{\hat{t}\hat{i}} \right) \gamma_{\hat{i}} \epsilon =0~,
}
which implies
\eqn\cbd{v_{\hat{t}\hat{i}} = {3\over 2} \partial_i U_1 e^{U_2}~.
}

The spatial part of the gravitino variation is \eqn\cbe{ \left[
\partial_i +{1\over 4} \omega_i^{~\hat{k}\hat{j}}
\gamma_{\hat{k}\hat{j}} + v^{\hat{t}\hat{j}}
\left(\gamma_{i\hat{t}\hat{j}} -{2\over3}
\gamma_i\gamma_{\hat{t}\hat{j}}\right) \right] \epsilon =0~. }
Substituting the formulae for the connection and using projection
\cba\ yields \eqn\cbf{ \left[ \partial_i + {1\over 2} \partial_j
U_2 \gamma_{\hat{i}\hat{j}} + v^{\hat{t}\hat{j}}
\left(\gamma_{i\hat{j}}-{2\over3}
\gamma_i\gamma_{\hat{j}}\right)\right]\epsilon =0~. } The radial
(${\hat i}={\hat j}$) and the angular (${\hat i}\neq{\hat j}$)
terms have a different form. Therefore they must vanish separately
so that
\eqn\cbh{\eqalign{
\left(\partial_i + {2\over 3}v^{\hat{t}\hat{i}}e_i^{~\hat{i}}\right)\epsilon &=0~,\cr
{1\over 2} \partial_j U_2  + {1\over3}v^{\hat{t}\hat{j}}e_i^{~\hat{i}}&=0~.
}}
Inserting $v_{{\hat t}{\hat i}}$ from \cbd\ into the second
equation we find that $U_1(x)=U_2(x)$. Therefore  we will drop the
index on the function $U(x)$ from now on. The first equation in
\cbh\ gives the form of the Killing spinor
\eqn\cbi{
\epsilon = e^{U(x)} \epsilon_0~,
}
where $\epsilon_0$ is a constant spinor.

In summary, the gravitino equation determines  the form of the
Killing spinor \cbi, the auxiliary field
\eqn\cbj{
v_{\hat{t}\hat{i}} = {3\over 2} \partial_i e^U~,~~~~v_{ti} = {3\over 2} e^{2U} \partial_i U~,
}
and simplifies the metric from \ca\ to
\eqn\cbk{ ds^2 = f^2 dt^2 - f^{-1} dx^i dx^i~, }
where $f=e^{2U(x)}$.

\vskip.2cm \noindent {\bf Gaugino variation} \vskip .2cm

We next analyze the gaugino variation, the second equation in \ab.
Noting that only $v_{\hat{t}\hat{i}}$ are non-vanishing, and using
the projection \cba, we have
\eqn\cca{
\left(-{1\over4}\gamma\cdot
F^I-{1\over2}\gamma^a\partial_aM^I-{2\over3}M^I v^{\hat{t}\hat{i}} \gamma_{\hat{i}}
\right)\epsilon=0~.
}
This requires that the scalars are time-independent  and that the
only non-zero components of the field strengths are
$F^{\hat{t}\hat{i}}$. This in turn gives the condition
\eqn\ccb{ -{1\over 2} F^{I\hat{t}\hat{i}}+{1\over 2}
e^i_{~\hat{i}}\partial_i M^I - {2\over3}M^I v^{\hat{t}\hat{i}} =0~.
}
Inserting the auxiliary field \cbj\ and switching to a  coordinate
frame we find
\eqn\ccc{
{1\over 2} e^{-2U} F^{I}_{it}-{1\over 2} \partial_i M^I  - M^I  \partial_i U =0~.
}
Reorganizing, we have
\eqn\ccd{
F^I_{it} = \partial_i (e^{2U} M^I )~,
}
which can be integrated to
\eqn\cce{
A^I_t = e^{2U} M^I~.
}
This equation captures the characteristic feature of attractor flows: the scalars follow
the electric potentials along the entire radial flow.

\vskip.2cm \noindent {\bf Auxiliary fermion variation} \vskip .2cm

Imposing $\delta\chi=0$ results in the condition
\eqn\cd{
\left(D-2\gamma^c\gamma^{ab}{\cal
D}_av_{bc}-2\gamma^a\epsilon_{abcde}v^{bc}v^{de}+
{4\over3}(\gamma\cdot v)^2\right)\epsilon=0~.
}
The third term vanishes in the spherically symmetric case and the fourth term can be evaluated as
\eqn\cda{
{4 \over 3} \left(\gamma\cdot v\right)^2= {16 \over 3} \delta^{ij} v_{\hat{t}\hat{i}} v_{\hat{t}\hat{j}}
= 12 e^{2U} \partial_i U \partial_i U ~.
}
For the covariant derivative we need the non-vanishing Christoffel symbols
\eqn\cdc{
\Gamma_{tt}^i = 2e^{6U} \partial_i U~,~~~~\Gamma_{ti}^t = 2 \partial_i U~,~~~~
\Gamma_{ij}^k = \left( \delta_{ij} \partial_k -\delta_{ik} \partial_j-\delta_{kj} \partial_i\right) U~.
}
and the auxiliary field \cbj. This gives the only non-vanishing
component of the covariant derivative as
\eqn\cdd{
{\cal D}_i v_{tj} = {3\over 2} e^{2U}\left( \partial_i\partial_j U + 2\partial_i U \partial_j U
- \delta_{ij} \partial_k U \partial_k U\right)~,
}
so that
\eqn\cde{
\gamma^c \gamma^{ab} {\cal D}_a v_{bc} = \gamma^i \gamma^{jt} {\cal D}_j v_{ti}
= -\gamma_{\hat t}\delta^{ij} {\cal D}_j v_{ti} = -{3\over 2} e^{2U}\gamma_{\hat t}
\left( \partial_i\partial_i U - 2\partial_i U \partial_i U\right)~.
}
After applying the projection \cba\ on the supersymmetry parameter $\epsilon$,
the condition \cd\ from the variation of the auxiliary fermion now becomes
\eqn\cdg{\eqalign{
D & = 3 e^{2U}
\left( \partial_i\partial_i U - 2\partial_i U \partial_i U\right) - 12 e^{2U} \partial_i U \partial_i U\cr
&= 3 e^{2U} \left( \nabla^2 U - 6 (\nabla U)^2\right)~.
}}

We have now exhausted the supersymmetry conditions \ab. As a result we have found
\cbj, \cce, and \cdg\ which determine $v_{ab}$, $A^I_t$, and $D$ in terms of the scalar
moduli $M^I$ and the metric function $U(x)$. These remaining functions are not
determined by supersymmetry alone. Instead we must now turn to the equations of
motion.

\subsec{Maxwell equations}

The black hole
solutions we seek are defined by conserved electric charges with respect to each gauge
field. For a given charge, Gauss' law determines the radial dependence of the electric field
as follows.

Neglecting the Chern-Simons terms, which do not contribute to spherically symmetric solutions,
the F-dependent terms in the Lagrangian are
\eqn\gza{\eqalign{
{\cal L}_F &=
2{\cal N}_I v^{ab}F^I_{ab} + {1\over 4} {\cal N}_{IJ} F^I_{ab} F^{Jab}\cr
&~~+{c_{2I}\over 24}\left( {1\over 6} F^I_{ab} v^{ab} D + {1\over 2} F^{Iab} C_{abcd} v^{cd}
-{4\over 3} F^{Iab} v_{ac} v^{cd} v_{db} - {1\over 3} F^{Iab} v_{ab} v^2\right)~.
}}
The Maxwell equations
\eqn\cec{ {\cal D}_\mu \left({\partial {\cal L} \over \partial
F^I_{\mu\nu}} \right)= {1\over\sqrt{g}}\partial_\mu \left(\sqrt{g}
{\partial {\cal L} \over \partial F^I_{\mu\nu}}\right)=0~, }
are equivalent to the statement $\p_r q_I =0$, where $q_I$ are the
conserved electric charges
\eqn\gzaa{ q_I  = - {1\over 4\pi^2 }\int_{S^3}  \sqrt{g}{\partial
{\cal L}\over\partial F^I_{tr}}
 = -{1\over 2} e^{-2U} r^3 {\cal E}^{r}_I~. }
The canonical momenta are
\eqn\gzb{ {\cal E}^{i}_I = {\partial {\cal L}\over \partial
F^I_{ti}} = 4{\cal N}_I v^{ti} + {\cal N}_{IJ} F^{Jti} +{c_{2I}\over
24} \left( {1\over 3} v^{ti} D + 2C^{titj} v_{tj} -{8\over 3} v^{tj}
v_{jt} v^{ti} - {2\over 3} v^{ti} v^2 \right)~. }
We need to make this expression more explicit. First, let us define
moduli with lower indices as\foot{This notation is actually redundant because $M_I={\cal N}_I$.}
\eqn\gzc{ M_I = {1\over 2} {\cal N}_{IJ} M^J~. }
In the
context of Calabi-Yau compactification of M-theory the $M_I$ are
volumes of four-cycles dual to the two-cycles with volume $M^I$.
At any rate, the definitions \aab\ of the various scalar functions now imply
\eqn\gze{ {\cal N}_{IJ} \partial_i M^J = \partial_i M_I~. }
We now find
\eqn\gzd{ 4{\cal N}_I v^{ti} + {\cal N}_{IJ} F^{Jti}=
e^{2U}\partial_i
 \left[ e^{-2U} M_I\right]~,
}
due to \cbj\ and \ccd.  It is straightforward to cast the
remaining terms in \gzb\ in this form as well, by using \cbj,
\ccd,\cdg, along with
\eqn\gzf{ C^{itjt} = - 2\partial_i\partial_j U - 6\partial_i U
\partial_j U +{3\over 2}\delta_{ij} (\nabla U)^2 + {1\over 2}
\delta_{ij} \nabla^2 U ~. }
After the dust has settled we find
\eqn\gzf{ {\cal E}^{i}_I  = e^{2U}  \partial_i\left[ e^{-2U} M_I -
{c_{2I}\over 8} (\nabla U)^2 \right]~. }
 We now see that the
Maxwell equation $\p_r q_I =0$ becomes
\eqn\gfa{
\nabla^2\left[ e^{-2U} M_I - {c_{2I}\over 8} (\nabla U)^2
\right]=0~, }
where $\nabla^2$ denotes the Laplacian on flat $\IR^4$.  Solutions are
thus a set of harmonic functions on this space. In this paper we
consider the single center solutions
\eqn\gfb{
e^{-2U} M_I - {c_{2I}\over 8} (\nabla U)^2
= H_I = M_I^\infty + {q_I\over r^2}~,
}
where the integration constants $M_I^\infty$ are the moduli at
infinity, and the $q_I$ are the same charges as appear in \gzaa.

\subsec{Completing the electric solution: the D equation} At this
point we have used supersymmetry to specify the entire solution in
terms of the functions $M^I$ and $U$, and we have determined $M^I$
by integrating Gauss' Law. Therefore, we need only one more
constraint to find the complete solution. For this we consider the
equation of motion for the auxiliary field $D$.

Starting from the Lagrangian ${\cal L}_0+{\cal L}_1$  given in
\ac\ and \af, the terms that depend on the $D$-field are
\eqn\ce{ {\cal L}_D = {1\over 2} ({\cal N} -1) D +
{1\over 24}c_{2I}\left( {1\over 12} M^I D^2 + {1\over
6}F^{Iab}v_{ab}D \right)~. }
Varying with respect to $D$ we find
\eqn\cea{ {\cal N} -1 + {c_{2I}\over 72} \left( M^I D +
F^{Iab} v_{ab}\right) =0~. }
 From  \cdg, \ccd, and \cbj, this becomes
\eqn\ceb{
{\cal N} -1 + {c_{2I}\over 24} e^{2U}\left(  (\nabla^2 U - 4 (\nabla U)^2) M^I
+ \nabla U \nabla M^I \right) =0~. }
In the absence of higher derivative terms this equation simply
reads ${\cal N}=1$. Since ${\cal N}={1\over 6} c_{IJK} M^I M^J
M^K$ this amounts to an algebraic constraint on the scalar
manifold. The  general equation with higher derivatives included
is much more complicated. To be explicit, recall that the $M^I$
are determined in terms of $M_I$ by the relation
\eqn\hca{
M_I ={1\over 2} c_{IJK} M^J M^K~,
}
and the $M_I$ in turn are given by
\eqn\hc{
M_I = e^{2U}\left( H_I + {c_{2I}\over 8}
(\nabla U)^2)\right)~,~~~~H_I = M_I^\infty + {q_I\over r^2}~.
}
The condition \ceb\ is thus a nonlinear, second order, ordinary
differential equation for $U(r)$.   But note that to write this
equation explicitly requires inverting \hca\ to find $M^I$, which
cannot be done until $c_{IJK}$ have been specified.\foot{ One also
should be alert to the fact that the inversion of \hca\ may not be
unique, which raises some interesting issues. Some explicit
examples in related contexts can be found in
\refs{\ShmakovaNZ,\ChouBA}.}   Once this has been done, the
resulting differential equation typically requires a numerical
treatment.

\subsec{Near Horizon Geometry} It is instructive to make the
equations above more explicit in the near horizon region of the
black hole. To do so take the integration constants $M_I^\infty=0$
and seek a solution with constant $M_I$ and $M^I$ related by \hca.
Then \hc\ gives $f=e^{2U} = {r^2\over\ell^2_S}$ and
\eqn\hxa{ \ell^2_S M_I = q_I + {1\over 8} c_{2I}~. }
The notation $\ell_S$ was chosen with some foresight. Indeed, the
change of variables $r^2 = \ell^3_S/2z$ bring the geometry \cbk\
into the standard form
\eqn\hxb{ ds^2 = {\ell^2_S\over 4z^2} (dt^2 -dz^2) - \ell^2_S
d\Omega^2_3~,}
which we recognize as AdS$_2\times S^3$ with $S^3$ radius $\ell_S$ and
AdS$_2$ radius $\ell_A = {1\over 2}\ell_S$.

In the near horizon region the $D$ equation \ceb\ is an algebraic
constraint
\eqn\hxc{ {\cal N} = {1\over 6} c_{IJK} M^I M^J M^K = 1+ {1\over
12\ell^2_S} c_{2I} M^I~. }
If we write the inversion equation \hca\ in terms of the rescaled variables
\eqn\hxd{ {\hat M}^I \equiv \ell_S M^I~, }
it becomes
\eqn\hxe{
{1\over 2} c_{IJK} {\hat M}^J {\hat M}^K = q_I + {1\over 8} c_{2I}~.
}
This is an algebraic equation that determines ${\hat M}^I$ as
functions of the charges $q_I$ and the numbers $c_{IJK}$ and
$c_{2I}$. Given such a solution, ${\hat M}^I={\hat M}^I(q_J)$, the
constraint \hxc\ gives the scale of the geometry
\eqn\hxf{
\ell^3_S = {1\over 6} c_{IJK} {\hat M}^I {\hat M}^J {\hat M}^K - {1\over 12} c_{2I} {\hat M}^I~,
}
where the right hand side is a function of the charges alone. Finally, \hxd\ and \hxe\
give the physical moduli in the near horizon region as
\eqn\hxg{\eqalign{
M_I & = {q_I + {1\over 8} c_{2I}\over \left( {1\over 6} c_{IJK} {\hat M}^I {\hat M}^J {\hat M}^K -
{1\over 12} c_{2I} {\hat M}^I
\right)^{2/3}}~, \cr
M^I & = {{\hat M}^I\over  \left( {1\over 6} c_{IJK} {\hat M}^I {\hat M}^J {\hat M}^K -
{1\over 12} c_{2I} {\hat M}^I
\right)^{1/3} }~.
}}
The expressions \hxe-\hxg\ completely specify the near horizon geometry of the 5D black
hole.

\subsec{Example: $K3\times T^2$}

In order to illustrate how our final expression \ceb\ determines
the entire radial dependence of the solution, we next
present a numerical solution in the special case of $K3\times
T^2$. As in the analogous magnetic example (section 3.5)
we let $M^1$ be the single modulus on the torus and
$M^i$ be the moduli of $K3$, where $i=2\ldots 23$.  The only
non-vanishing intersection numbers are then $c_{1ij} \equiv
c_{ij}$, where $c_{ij}$ is the intersection matrix for $K3$, with
inverse $c^{ij}$.

 From $M_I = \half c_{IJK} M^J M^K$, we find
\eqn\da{ M_1 =\half
c_{ij} M^i M^j~, ~~~~~M_i = c_{ij}  M^j M^1~, }
We easily invert this to obtain the $M^I$ as functions of the
$M_I$
 \eqn\dc{ M^1 = \sqrt{c^{ij} M_i M_j \over 2 M_1}~, ~~~~~M^i
= c^{ij} M_j \sqrt{ 2M_1 \over c^{kl} M_k M_l }~. }

The Chern class $c_{2I}$ is calculated on the  4-cycle Poincare
dual to the $I$-th 2-cycle. Therefore, the $c_{2i}$ vanish,
leaving only $c_{2,1}=c_2(K3)=24$.

Substituting \hc\ into \dc\ to yields
\eqn\de{ M^1 = \left( {e^{2U}
c^{ij} H_i H_j \over 2H_1 +  6   (U^\prime)^2 }
\right)^{1/2}~, ~~~~~M^i = \left({e^{2U}c^{ij}H_i H_j \over 2 H_1
+ 6 (U^\prime)^2}\right)^{-1/2} e^{2U}c^{ij}H_j~, }
where primes denote derivatives with respect to $r$. The special geometry constraint \ceb\ is
\eqn\df{ \half c_{ij}M^i M^j M^1 -1 + e^{2U}\left[(
U^{\prime\prime}+{3\over r}U^\prime - 4U^{\prime 2} )M^1  + U^\prime
(M^1)^\prime\right] =0~. }
The problem is now to insert \de\ into \df\ and solve for $U(r)$.

This is straightforward to solve numerically, given specific choices
of charges. Consider a small black hole, $q_1=0$ with $q_2= q_3 =1$,
$c^{23}=1$. We also assume $H=H_2=H_3=1+{1\over r^2}$ are the only
harmonic functions not equal to unity. Then \df\ becomes
\eqn\dgd{H U''+(1+3(U')^2)\left[U'r^{-1}\left(3+{1\over
r^2}\right)+H\right] -e^{-3U}(1+3(U')^2)^{3/2}=0~.}
The boundary conditions are fixed by matching to the desired small
$r$ behavior
\eqn\ceq{e^{-2U} \sim {\ell_S^2 \over r^2}~,}
with $\ell_S=3^{-1/6}$. The result of the numerical analysis is the $U(r)$ shown in Figure 2.
It exhibits the same kind of oscillations seen in \refs{\DabholkarDQ,\senstretch}
and discussed in the end of section 3.5.
\fig{Numerical solution of \dgd. The plot displays $e^{-2U(r)}$ for
small $r$.}{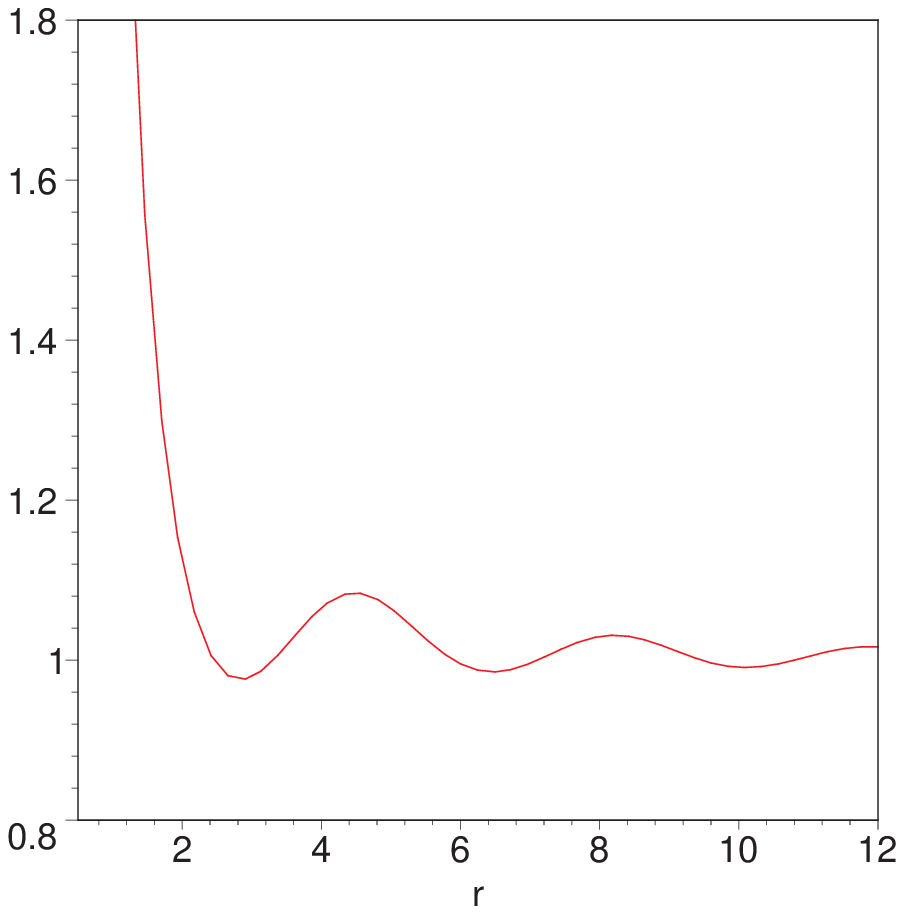}{3.0truein}

\newsec{Entropy of electrically charged black holes}

\subsec{Entropy function} We now turn to the computation of the
entropy of the electric black hole solutions. Due to the higher
derivative corrections, the relevant formula is Wald's generalized
expression for the entropy as a surface integral over the horizon
\wald.  Generally  it can be laborious to integrate Wald's entropy
density, but the extremal black holes we consider have a near
horizon AdS$_2$ factor, and for such black holes the problem
reduces to evaluating an ``entropy function" at its extremum
\SenWA. The entropy function is the Legendre transformation of the
action with respect to the electric charges. In our conventions,
the precise expression is
\eqn\za{S= \pi \ell_A^2 \ell_S^3 ( F^I_{tr}{\p {\cal L} \over\p
F^I_{tr} } -{\cal L})~,}
where ${\cal L}$ is the Lagrangian density evaluated in the near
horizon AdS$_2 \times S^3$ geometry.

\subsec{Near horizon supersymmetry} To proceed we  need to
evaluate ${\cal L}={\cal L}_0 +{\cal L}_1$, and its derivative
with respect to the field strength. This task is greatly
simplified by taking advantage of the conditions resulting from
enhancement of supersymmetry in the near horizon region. This
means there is no need to impose any projector condition on the
Killing spinor $\epsilon$, and so each term in the supersymmetry
conditions \ab\ vanishes by itself, rather than balancing off
other terms.

To see how this  works, recall that the variation of the auxiliary
fermion includes the term \cbe
\eqn\zb{
\gamma^c \gamma^{ab} {\cal D}_a v_{bc} = -{3\over 2} e^{2U}\gamma_{\hat t}
\left( {1\over r^3}\partial_r (r^3\p_r U) - 2(\partial_r U)^2 \right)~.
}
Taking this to vanish we immediately find a metric function of the
form
\eqn\zc{e^{2U}={r^2 \over \ell_S^2}~,}
for some $\ell_S$. This is recognized as the metric function describing
AdS$_2\times S^3$ \hxb\ with the scales of the constituent spaces
related as
\eqn\zd{\ell_A = \half \ell_S~.}

With the geometry in hand, the $D$-field \cdg\
determined from the variation of the auxiliary fermion
becomes
\eqn\zda{
D = 3e^{2U} \left( \partial^2 U - 6(\partial U)^2\right)
 = -{12\over\ell^2_S}~,
}
and the
auxiliary two-form \cbj\ determined from the
gravitino variation becomes
\eqn\ze{ v_{tr} = {3\over 2}e^{2U}\partial_r U = {3r\over
2\ell^2_S}~. }
The gaugino variation \cca\ shows that  the moduli $M^I$ are
constants in the near horizon region. It also gives the field
strengths \ccd\ as
\eqn\zf{ F^I_{tr} = -\partial_r(e^{2U}M^I) = -{2r \over \ell_S^2}M^I~.}
Our general expression \gzaa\ for the electric charge can be written in
the near horizon region as
\eqn\zg{ q_I
= -\half \ell_S^2 r {\p {\cal L} \over \p F^I_{tr}
}~.}
Using \zd, \zf, and  \zg, the entropy function becomes
\eqn\zga{ S = \pi(\ell_S q\cdot M  -{\ell_S^5 \over 4}{\cal L})~.}

\subsec{Evaluation of entropy function}
Up until now we have just used the supersymmetry variations, which
are independent of the action.  We now need to use details of the
action.  The first piece of information we need is the modified
special geometry constraint \ceb\
coming from the $D$ equation of motion.
Using the near horizon field values found above we recover \hxc\
\eqn\zk{ \Nc -1 -{c_2 \cdot M \over 12 \ell_S^2}=0~.}

Next, we need to evaluate the Lagrangian density.  Using the near
horizon supersymmetry results, as well as \zk, we find for the two
derivative Lagrangian ${\cal L}_0$,
\eqn\zl{{\cal L}_0 = {4 \over \ell_S^2} +{1 \over 3} {c_2 \cdot M
\over \ell_S^4}~,}
after some algebra. For the four derivative Lagrangian ${\cal L}_1$,
\eqn\zl{{\cal L}_1 = -{1 \over 2} {c_2 \cdot M \over \ell_S^4}~,}
after more algebra. Altogether
\eqn\zm{ {\cal L} = {\cal L}_0 +{\cal L}_1 = {4 \over \ell_S^2}
-{1 \over 6} {c_2 \cdot M \over \ell_S^4}~,}
giving the entropy function
\eqn\zn{ S= \pi  (\ell_Sq\cdot M - \ell_S^3 +{1 \over 24}\ell_S
c_2 \cdot M)~.}

At  this stage we could evaluate $S$ by inserting the values for
$\ell_S$ and $M^I$ obtained from our explicit solutions, but it is
more instructive to proceed by extremizing the entropy function.
This also serves as a useful consistency check on our results.

The problem consists of extremizing $S$ with respect to $\ell_S$
and $M^I$, while holding fixed $q_I$ and imposing the constraint
\zk.   We therefore  add in a Lagrange multiplier and write
\eqn\ia{ S = \pi  \left( \ell_S q\cdot M  - \ell_S^3 +{1 \over 24}
 \ell_S c_2 \cdot M + \lambda(\Nc-1- {c_2 \cdot M \over 12 \ell_S^2})
\right)~.}
Extremizing gives
\eqn\ib{\eqalign{ 0& = q \cdot M -3 \ell_S^2 +{1 \over 24} c_2
\cdot M +{1 \over 6} \lambda {c_2 \cdot M \over \ell_S^3}~, \cr
0& =\ell_S q_I +{1 \over 24} \ell_S c_{2I}+ \lambda \Nc_I  -{1 \over
12 }\lambda {c_{2I} \over \ell_S^2}  ~,\cr
0&= \Nc-1 - {c_2 \cdot M\over 12 \ell_S^2}~. }}

We can solve for $\lambda$ as follows.  Contract the second
equation with ${1 \over \ell_S}M^I$, subtract it from the first,
and use the third to eliminate $\Nc$.  This gives
\eqn\ibc{\lambda =- \ell_S^3~.}

Before continuing, we can use these equations to rewrite the
entropy in a suggestive form.  Using \ibc, and the first and third
equations of \ib, we  insert into \ia\ to get
\eqn\ibz{ S=  2\pi \Nc \ell_S^3~. }
This is the same formula as we would find in the two derivative
theory, except that in that case we would have $\Nc=1$.  From the
higher dimensional point of view the condition $\Nc=1$ corresponds
to fixing the $CY_3$ to have unit volume.  More generally, if we
continue to think of $\Nc$ as the volume, then we see that \ibz\
is precisely $S = A/4G$, with $A$ being the horizon area in $11$
dimensions.  However, one should perhaps not take this too
seriously, since in the presence of higher derivatives the metric,
and hence the horizon area, are subject to field redefinition
ambiguities.

 From \ibc\ and the middle equation of \ib\  we can now solve for
$\Nc_I$ as
\eqn\ibd{\ell_S^2 \Nc_I = q_I +{1 \over 8}c_{2I}~. }
This agrees with our previous result \hxa, which we obtained by
integrating Gauss' law and matching onto charges defined in the
asymptotically flat region. The agreement is a nontrivial check
on the consistency of our method (and the accuracy of our algebra).

We are now ready to find the entropy. Introducing the rescaled
moduli
\eqn\ibzb{ {\Mh}^I = \ell_SM^I ~,}
as in \hxd, the entropy \ibz\ becomes simply
\eqn\ibzc{ S =2\pi \left({1 \over 6} c_{IJK}\Mh^I \Mh^J \Mh^K
\right)~. }
The rescaled moduli can be found by solving \ibd\ written
in the form
\eqn\ibzd{ \half c_{IJK} \Mh^J \Mh^K =\qh_I~,}
where the shifted charge is defined as
\eqn\ibza{\qh_I = q_I +{1 \over 8} c_{2I}~.}
The solution to \ibzd\ will take the form $\Mh^I(\qh_I)$ which we then insert
in \ibzc\ to find the entropy as function of the charges.

The value of $\ell_S$ can be computed by solving the special
geometry constraint \zk\ from which we recover our previous result
\hxf\ for $\ell_S$.  However, we do not actually need $\ell_S$ to
find the entropy, because the factors of $\ell_S$ were scaled away
when arriving at the entropy formula \ibzc.

The computation of the entropy in terms of the $\qh_I$ is almost
insensitive to the detailed form of the action. All we need is \ibzc\
and \ibzd\ which could be derived using just the conditions
due to enhancement of supersymmetry.
To get the right shift in the definition \ibza\ of $\qh_I$, though,
we need to use some information about the action,
such as the $D$ equation of motion. Assuming we know this shift,
we see that if we know the black hole entropy in the two
derivative theory, then the corrected entropy is obtained simply
by replacing the charges by the shifted charges.

Strictly speaking, the regime of validity of our computation only
extends to terms first order in $c_{2I}$, since we only considered
the addition of four-derivative terms to the action.  {\it A
priori}, $2+2n$ derivative terms in the action will contribute at
the same order as any $(c_{2I})^n$ terms.   In the case of black
holes / strings with near horizon geometry AdS$_3 \times S^2$, one
can use anomalies and supersymmetry to prove that the four
derivative action in fact gives the exact expression for the large
momentum behavior of the entropy \refs{\us,\KrausZM}.  In the
present case it is also tempting to conjecture that the regime of
validity extends beyond the first order terms, at least in some
cases.

\subsec{Comparison with other results}

\vskip.2cm \noindent {\bf Comparison with 4D black holes and the
topological string} \vskip .2cm

In \GuicaIG\ an entropy formula for 5D black holes was
conjectured, based on 4D results and the topological string
partition function.   The entropy is given by
\eqn\yya{S= {\cal F} - \phi^I {\partial {\cal
F}\over\partial\phi^I}~, }
with
\eqn\yyb{ {\cal F} = -{1\over\pi^2} \left( D_{IJK} \phi^I \phi^J
\phi^K - {\pi^2\over 6}c_{2I} \phi^I\right)~. }
This yields
\eqn\yyc{S= {2 \over \pi^2}D_{IJK}\phi^I \phi^J \phi^K~.}%
To convert to our notation, use
\eqn\yyd{D_{IJK}={1 \over 6}c_{IJK}~,\quad \phi^I = \pi
\hat{M}^I~,}
so that the entropy becomes
\eqn\yye{ S =2\pi \left({1 \over 6} c_{IJK}\Mh^I \Mh^J \Mh^K
\right)~, }
in precise agreement with \ibzc.

On the other hand, in \GuicaIG\ the electric charges are
\eqn\yyf{ q_I = -{\p{\cal F}\over \p \phi^I} = {1 \over 2} c_{IJK}
\Mh^J \Mh^J -{1 \over 6} c_{2I}~,}
which is equivalent to
\eqn\yyg{ \qh_I = q_I+{1\over 6} c_{2I}~.}
This is to be compared with \ibza.  Thus, when expressed in terms
of the $q_I$ our entropy formula does not agree with \GuicaIG. The
reason for this mismatch is explained in detail in \CastroCI.  The
shift \yyg\ gives the wrong result for the 5D black hole since it
includes an extra charge induced by the Taub-NUT geometry.

We also note that the authors of \GuicaIG\ performed a 5D
supergravity computation keeping only the Gauss-Bonnet like term.
This computation yielded a different discrepancy, which is not
surprising since the full action contains many more terms at this
order.

\vskip.2cm \noindent {\bf  $K3 \times T^2$ black holes} \vskip
.2cm

In general we have to invert \ibzd\ to express the entropy in
terms of the electric charges $q_I$.   This can be done explicitly
when $c_{IJK}$ are the intersection numbers for $K3\times T^2$.
This is basically the same problem we solved in \dc, with solution
\eqn\ibzg{\Mh^1 = \sqrt{ c^{ij}\qh_i \qh_j \over 2 \qh_1}~,\quad
\Mh^i=\sqrt{ 2 \qh_1 \over  c^{kl}\qh_k \qh_l } c^{ij}q_j~. }
 The entropy is then
\eqn\ibzh{\eqalign{S &=\pi \sqrt{ 2\qh_1 c^{ij}\qh_i \qh_j} =\pi
\sqrt{ 2(q_1 + 3)c^{ij}q_i q_j} ~.}}
A small black hole corresponds to taking $q_1=0$, such that the
M2-branes lie entirely within $K3$.   The higher derivative terms
give a finite size horizon to this would-be singular charge
configuration.

\vskip.2cm \noindent {\bf Elliptically fibred Calabi-Yau black
holes} \vskip .2cm

There is no known microscopic description of  black holes made
from wrapping M2-branes on a generic Calabi-Yau.  However, in
\VafaGR\ Vafa proposed such a description for an elliptically
fibred Calabi-Yau.  This proposal yields a correction to the
entropy that has the right form to match with a four derivative
term in five dimensions.  We can use our results to check that the
coefficient also agrees.

Consider M-theory on a Calabi-Yau 3-fold $K$ realized as an
elliptic fiber $E$ over a base space $B$. Wrap $M2$ branes along a
two-cycle $[C]+n[E]$, {\it i.e.} one that has components along the
fiber and also along $C\subset B$.  In \VafaGR\ it was argued that
the relevant moduli space is the symmetric product ${\rm
Sym}^n(\hat{C})$, where $\hat{C}$ is an elliptically fibred four
manifold with base $C$.  BPS states are then computed from the
cohomology of $\hat{C}$ in the standard fashion. The cohomology
leads to the entropy formula
\eqn\zza{S = \pi\sqrt{2 n(C\cdot C + 3 c_1(C) +2)}~,}
where the intersection products refer to the base $B$.  This
formula is valid for large $n$.  Now, if we follow \VafaGR\ and
use $c_1(C) = {1 \over 12} c_2(\hat{C})$, we see that the leading
order correction to the entropy corresponds to the shift
\eqn\zzb{ C\cdot C \rightarrow C\cdot C +{1 \over 4}
c_2(\hat{C})~.}
This matches the leading order shift obtained from \ibza,
\eqn\zzc{ \qh \cdot \qh \approx q \cdot q +{1 \over 4} c_2 \cdot
q~.}

\bigskip
\noindent {\bf Acknowledgments:} \medskip \noindent   The work of
PK and JD is supported in part by NSF grant PHY-0456200. The work
of FL and AC is supported  by DOE under grant DE-FG02-95ER40899.

\appendix{A}{Conventions}

 We briefly summarize our conventions.
 The metric signature is mostly minus $\eta_{ab}= {\rm
diag}(+,-,-,-,-).$\foot{The signature is opposite to that in our
previous paper \CastroSD.} Covariant derivatives of spinors are
defined as
\eqn\ai{ {\cal D}_\mu =\partial_\mu + {1\over4} \omega_\mu^{~ab}
\gamma_{ab}~, }
where $\omega^{ab}$ are the spin-connection one forms related to
the vielbein through the Cartan equation
\eqn\aia{ d e^a  +  \omega^a_{~b} \wedge e^b  =0~. }
Our convention for the curvature is
\eqn\ahb{ R^\lambda_{~\mu\nu\kappa} = \partial_\kappa
\Gamma^\lambda_{\mu\nu}-\partial_\nu \Gamma^\lambda_{\mu\kappa}
+\Gamma^\sigma_{\mu\nu} \Gamma^\lambda_{\kappa\sigma}
-\Gamma^\sigma_{\mu\kappa} \Gamma^\lambda_{\nu\sigma}~.}
The scalar curvature is then, e.g., $R={p(p-1) \over
\ell_A^2}-{q(q-1) \over \ell_S^2}$ for AdS$_p\times S^q$.  The
Weyl tensor is given by
\eqn\ahc{ C_{abcd}=R_{abcd}- {2 \over
3}(g_{a[c}R_{d]b}-g_{b[c}R_{d]a})+{1 \over 6}g_{a[c}g_{d]b}R~. }
Anti-symmetric products of gamma-matrices are normalized so that
$\gamma_{abcde} = \varepsilon_{abcde}$ where
$\varepsilon_{01234}=1$.

Finally, we take $G_5={\pi\over 4}$ and measure moduli in units of
$2\pi\ell_{11}$. In these units the charges are quantized (for
review see \LarsenXM).

\listrefs
\end